\documentclass[journal]{IEEEtran}
\usepackage[english]{babel}
\usepackage{nicematrix}
\usepackage[top=2cm,bottom=2cm,left=1.5cm,right=1.5cm,marginparwidth=1.75cm]{geometry}
\usepackage{graphicx,subfigure,mhchem,amsmath,cite}
\usepackage[colorlinks=true, allcolors=blue]{hyperref}

\title{Synchronization of molecular electrochemical oscillators by photon-assisted entanglement}

\author{Serge Kernbach \\[3mm]
\small CYBRES GmbH, Research Center of Advanced Robotics and Environmental Science,\\
\small Melunerstr. 40, 70569 Stuttgart, Germany, {\it serge.kernbach@cybertronica.de.com}
\vspace{-7mm}
}

\begin{document}
\maketitle

\begin{abstract}
Formation of hydronium and carbonate ions from carbon dioxide in the aqueous phase is a reversible process and can both produce and consume ions. These equilibrium reactions represent molecular electrochemical oscillators in pure water. Reversible switching of ionic dynamics is a chaotic process, which is influenced by the \ce{CO_2} level, temperature, concentration of decay products, pressure, magnetic fields and other factors. As demonstrated in previous works, para- and ortho- isomers of water have different electrochemical reactivity; weak variations of magnetic fields induce a low-energy spin conversion process between isomers and affect several electrochemical and physical parameters. In particular, it is expected that  spin-controlled ionic reactivity on different time scales can lead to macroscopic synchronization effects in the dynamics of microscopic electrochemical oscillators. This work explores this hypothesis by monitoring the high-resolution ionic dynamics and temperature of independent fluidic cells with electrochemical impedance spectroscopy. The occurrence of synchronization is studied in 4-16 cells grouped in one or several non-transparent thermo-insulating containers; about 20 million of samples are analyzed. Synchronization effects are shown to occur primarily in the \ce{CO_2} dissolving scenario on the 3-10 minute scale. Without \ce{CO_2} access, mutual synchronization is either non-existent or negligible. Maximal correlations with $r>0.9$ are achieved between 4-6 cells with one synchronization event per 8000 samples; with $r>0.7$ -- in up to 8-10 cells with one event per 3000 samples. Anti-phase correlations occur more frequently than in-phase correlations in all setups. The number of synchronization events is about five times lower when cells are separated between non-transparent containers. We also noted a generation of in-phase and anti-phase temperature-impedance waves highly synchronized between independent cells. To explain such results, we consider molecular quantum networks that operate with spin conversion of water isomers. Weak coupling between oscillators in independent cells can be introduced by photon-assisted entanglement triggered by slight variations of magnetic fields.
\end{abstract}

\section{Introduction}

Coupled nonlinear oscillators are well known in theoretical physics \cite{Atmanspachera05}, physical chemistry \cite{doi:10.1063/1.1426382}, adaptive control~\cite{Konishi99} and other disciplines, where different spatio-temporal effects are observed \cite{Chate92}. Coupled quantum oscillators are used to study macroscopic entanglement in micro-mechanical systems \cite{doi:10.1126/science.abf2998}, distant spin arrays \cite{thomas2021entanglement}, optomechanics \cite{wang2016macroscopic} and other systems. Oscillating reactions are also known in electrochemistry, e.g. spontaneous oscillations of electrode potential \cite{chan2019spontaneous}, the Belousov–Zhabotinsky, Bray–Liebhafsky, Briggs–Rauscher reactions \cite{ullah2019determination, bai2017periodic} or biochemical peroxidase–oxidase and liquid membrane oscillators \cite{chan2019spontaneous, peng2017electrochemical}. The formation of hydronium and carbonate ions \ce{H_3O^+}, \ce{HCO_3^-} and \ce{CO_3^{2-}} from carbon dioxide in the aqueous phase can also be considered from the viewpoint of coupled oscillators.

These processes are of interest since the ionic productivity is described by equilibrium reactions with forwards and reverse phases. Switching between phases generates a chaotic electrochemical dynamics that is detectable by electrochemical impedance spectroscopy (EIS) \cite{Kernbach17water}. Recent publications present several mechanisms that can trigger forward/reverse phases, such as ionic saturation \cite{Capobianco14}, temperature \cite{Mitchell09} or weak excitation of water by light or EM fields \cite{Vaskina2020}. The excited water is considered to be in a significantly non-equilibrium state with respect to the spin temperature, tending to an equilibrium state after the excitation is removed \cite{Pershin09Temp}. Independent measurements confirmed different ionic reactivities \cite{kernbach2022electrochemical}, surface tension \cite{kernbach2023Pershin}, heat capacity \cite{kernbach23Thermal} and evaporation \cite{POULOSE2023814} triggered by low-energy excitations at $10^{-8}$J/mL. These results are considered to be related to a spin-conversion process between para- and ortho-isomers in ice-like structures on different time scales \cite{Monserrat20}, \cite{Pershin15Biophysics}.

This work considers switching between forward/reverse phases of ionic production as molecular oscillators influenced by above mentioned processes. Weak coupling between oscillators is represented by molecular mechanisms, e.g. exchange of protons in hydrogen-bound networks \cite{Konyukhov11} or by light-matter interactions \cite{Dovzhenko18}. Denoted as photon-assisted entanglement \cite{Zhang20}, this mechanism is found in different microscopic physical \cite{kotler2021direct} and biological \cite{Marletto_2018} systems, and is responsible for emergent behavior on macroscopic scales. Following this approach, it is expected to detect such synchronization effects in electrochemical systems that are observed in coupled quantum oscillators \cite{Frimmer17, Choi21},  spin ensembles under light excitation \cite{TANG2022} or opto-micromechanics \cite{Korppi18}. Since spin-level phenomena are involved into macroscopic synchronization, investigation of these quantum mechanisms can be conducted with low-cost EIS or mobile NMR \cite{10.3389/fpls.2021.617768, Pershin09NMR} sensors, which can lead to new quantum sensing technologies.

This work experimentally explores this hypothesis in setups already published in \cite{kernbach2022electrochemical}. \ce{CO_2} dissolving is monitored by EIS in 4-16 optically transparent electrochemical cells with pure water placed in one or several thermo-insulating containers. Control attempts include optically non-transparent setups, electrochemical processes without \ce{CO_2} dissolving as well as different temperature-dependent phenomena. Additionally, various environmental parameters are recorded to test correlations between e.g. mechanical or electromagnetic impacts and synchronized electrochemical dynamics.

\section{Coupled electrochemical oscillators}

\begin{figure*}
\centering
\subfigure{\includegraphics[width=0.7\textwidth]{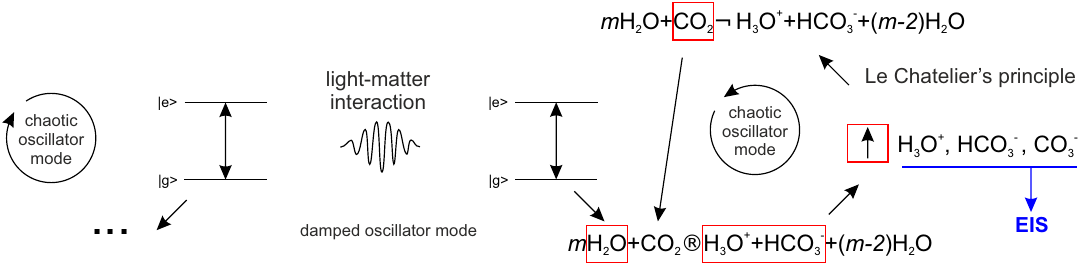}}
\caption{Proposed scheme of coupled electrochemical oscillators.  \label{fig:oscillatorScheme}}
\end{figure*}

Proposed scheme of coupled electrochemical oscillators is shown in Fig. \ref{fig:oscillatorScheme}.
\ce{CO_2} dissolving in water forms the carbonic acid \ce{H_2CO_3}
\begin{equation}
\label{eq:carbonicAcid}
H_2O + CO_{2(aq)} \rightleftharpoons H_2CO_3
\end{equation}
that dissociates to \ce{H_3O^+} and \ce{HCO_3^-}. Following \cite{Capobianco14}, the ionic dynamics of
\begin{equation}
\label{eq:carbonicAcidIons}
mH_2O + CO_2 \rightleftharpoons H_3O^+ + HCO_3^- + (m-2)H_2O
\end{equation}
is monotonic with short-term oscillation in initial phase. Bicarbonate $HCO_3^-$ further dissociates and forms \ce{CO_3^{2-}}
\begin{equation}
\label{eq:carbonicAcidIonsFurther}
HCO_3^- \rightleftharpoons  H_3O^+ + CO_3^{2-}.
\end{equation}
Dynamics of \ce{H_3O^+}, \ce{HCO_3^-}, \ce{CO_3^{2-}} production is investigated in \cite{Mitchell09}. The equilibrium reactions (\ref{eq:carbonicAcid}), (\ref{eq:carbonicAcidIons}), (\ref{eq:carbonicAcidIonsFurther}) are obeying Le Chatelier's principle; they have forward and inverse directions that increase or decrease ionic content. As demonstrated in \cite{kernbach2022electrochemical}, magnetic field and light excitations change the productivity (reactivity) of (\ref{eq:carbonicAcid}) that affects the equilibrium conditions. Increase of carbonic acid leads to more ions in fluids, enrichment by ions triggers the inverse reactions. In this way we observe chaotic electrochemical dynamics explained by multiple uncoordinated reactions (\ref{eq:carbonicAcid})-(\ref{eq:carbonicAcidIonsFurther}) in bulk water. Since para- and ortho- isomers of water have different electrochemical reactivity \cite{Kilaj18}, a spin conversion between isomers is considered to be one of mechanisms affecting equilibrium conditions of (\ref{eq:carbonicAcid})-(\ref{eq:carbonicAcidIonsFurther}). Recent publications \cite{Pershin09Temp} argue that spin conversion takes place in ice-like structures of interface water with a long lifetime of non-equilibrium ratio between isomers \cite{Monserrat20}. However, even short-term fluctuations between isomers affect the productivity of (\ref{eq:carbonicAcid})-(\ref{eq:carbonicAcidIonsFurther}); thus, isomers with both short and long lifetimes are involved in the chaotic electrochemical dynamics. In parallel to \ce{CO_2} dissolving, \ce{H_2O_2} in ROS reactions produces similar chaotic electrochemical dynamics, related to excited spin stage of singlet oxygen \cite{kernbach2022electrochemical}. 

\begin{figure}
\centering
\subfigure[\label{fig:lightIsolatedSetupResult}]{\includegraphics[width=0.495\textwidth]{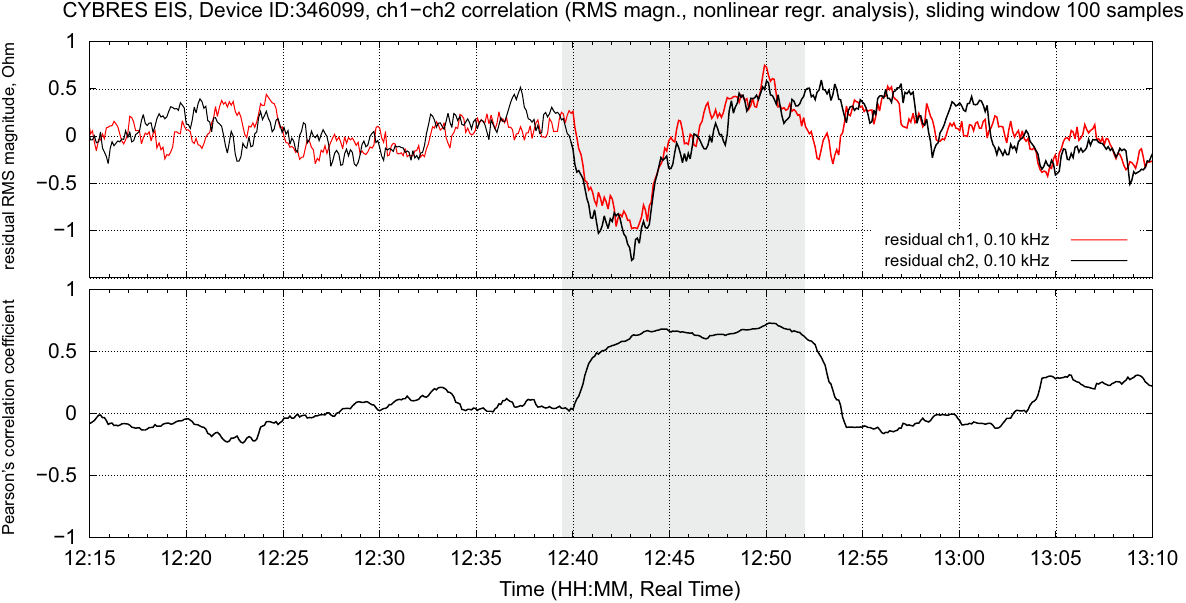}}
\subfigure[\label{fig:shortTermCorr}]{\includegraphics[width=0.49\textwidth]{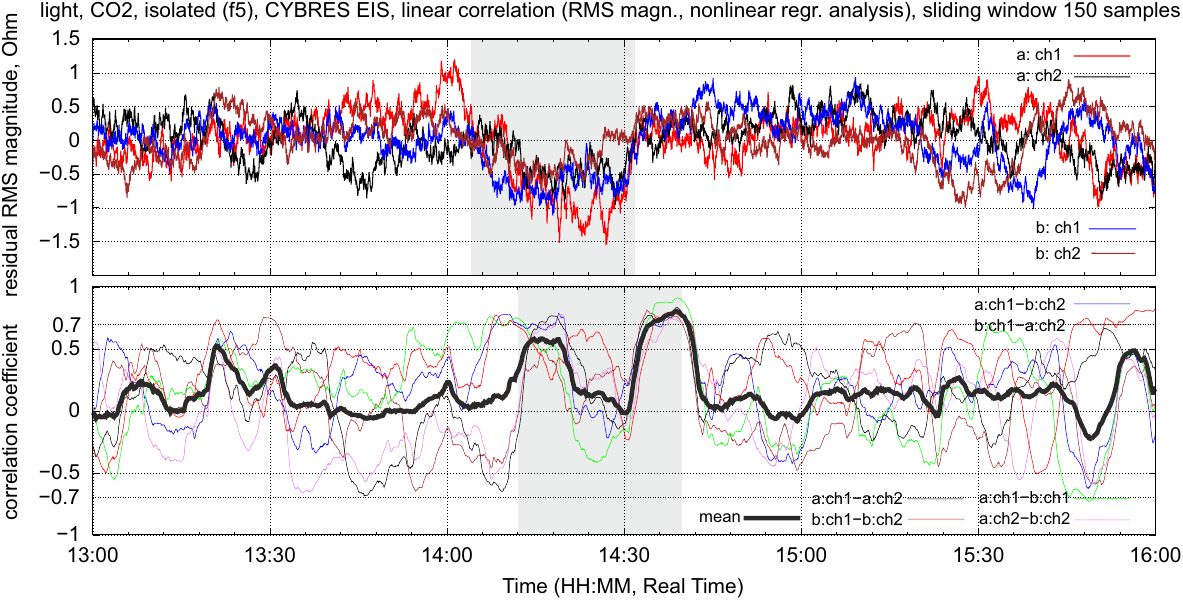}}
\caption{\textbf{(a)} Example of a correlated dynamics of electrochemical impedances from two fluidic cells. Dynamics of environmental parameters for this experiment is shown in Fig. \ref{fig:shortTermCorrAdd}, we do not observe any correlations with electrochemical impedances; \textbf{(b)} Example of EIS dynamics with 4 cells and six correlation coefficients. Polynomial (\ref{eq:approx2}) of 5$^{th}$ order is used for the analysis. \label{fig:singleWave}}
\end{figure}

\begin{figure*}[htp]
\centering
\subfigure[\label{fig:experimentalSetup}]{\includegraphics[width=0.625\textwidth]{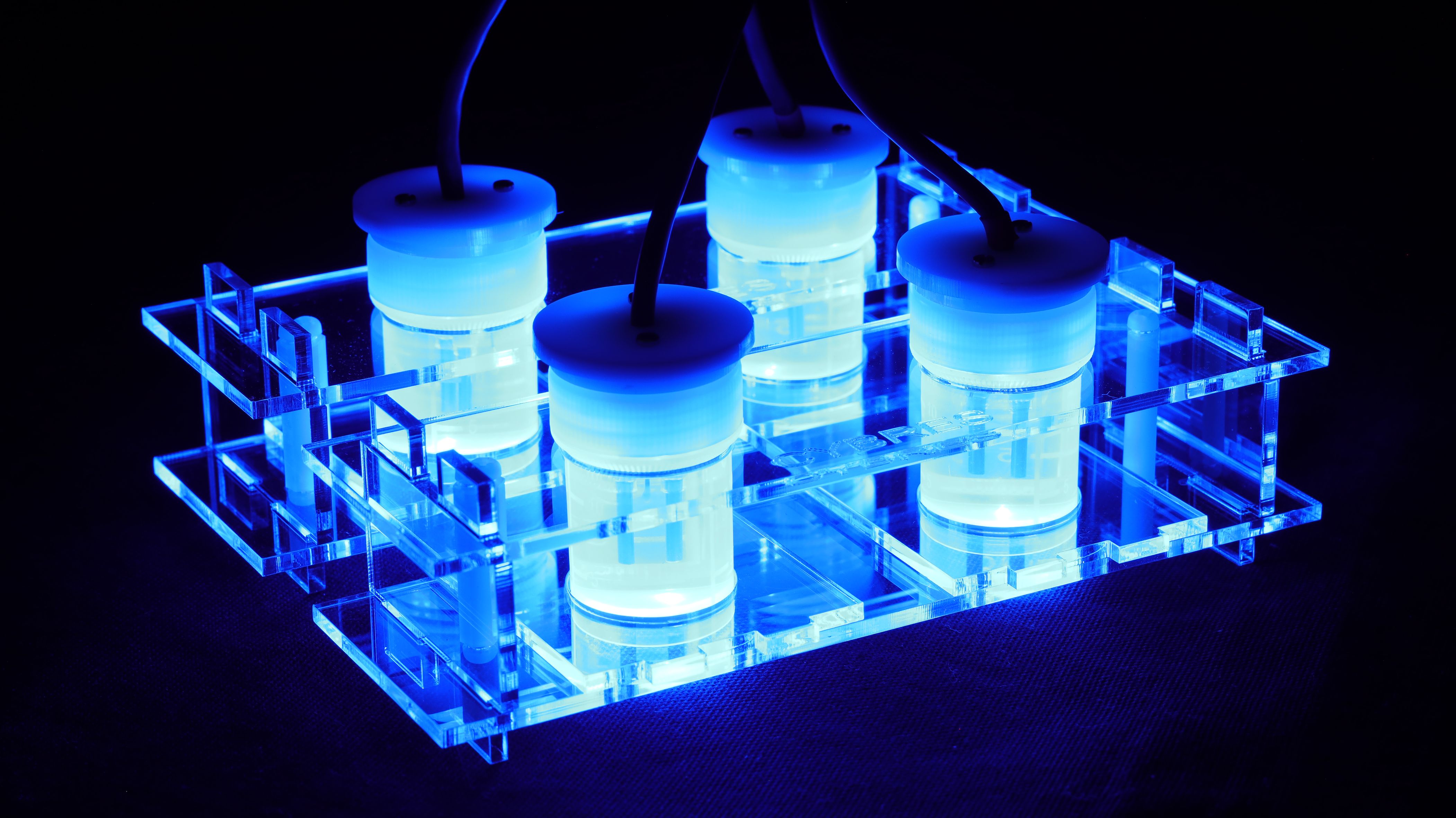}}~
\subfigure[\label{fig:CO2_scenario_A}]{\includegraphics[width=0.36\textwidth]{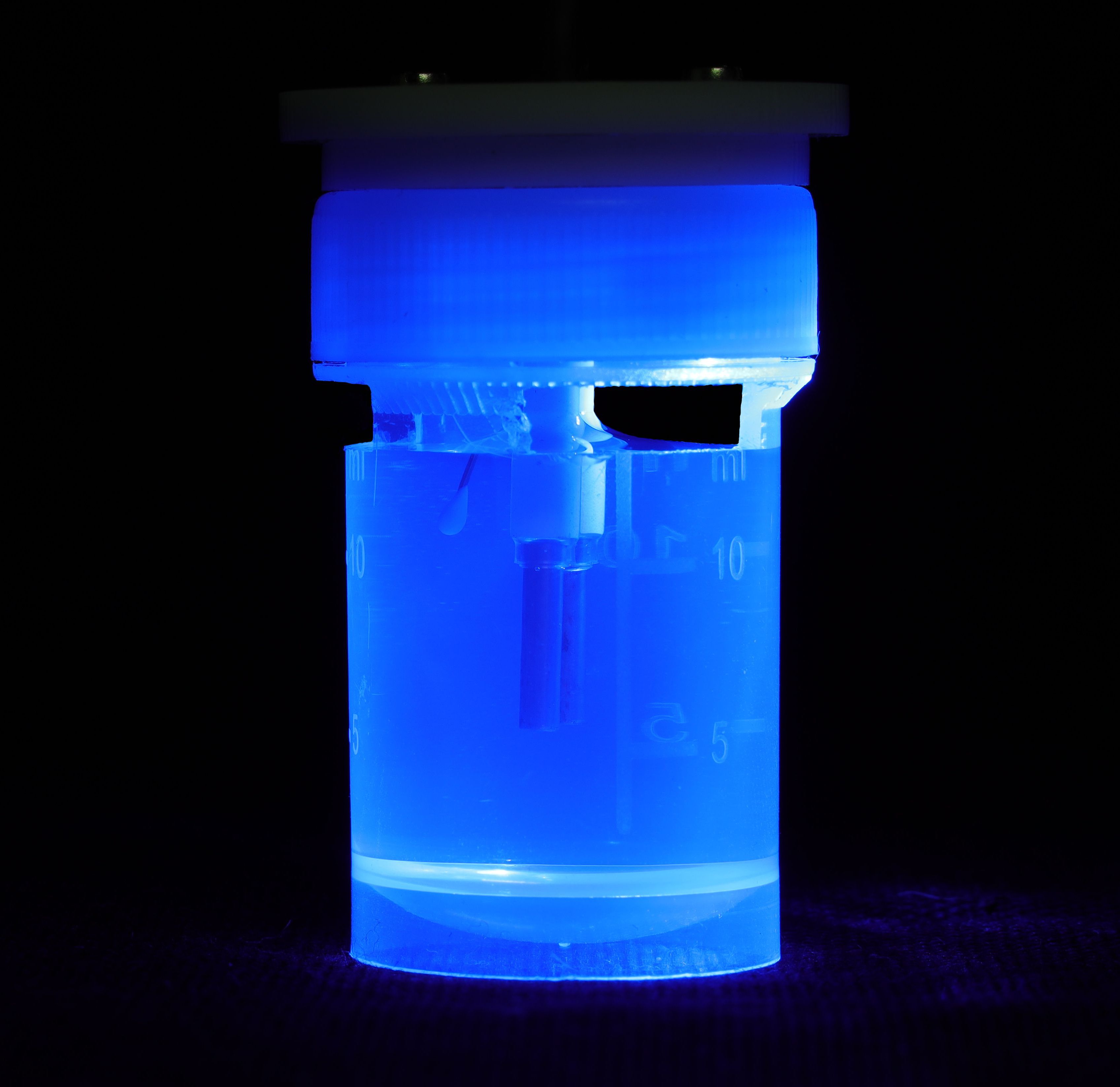}}\\
\subfigure[\label{fig:CO2_scenario_B}]{\includegraphics[width=1.0\textwidth]{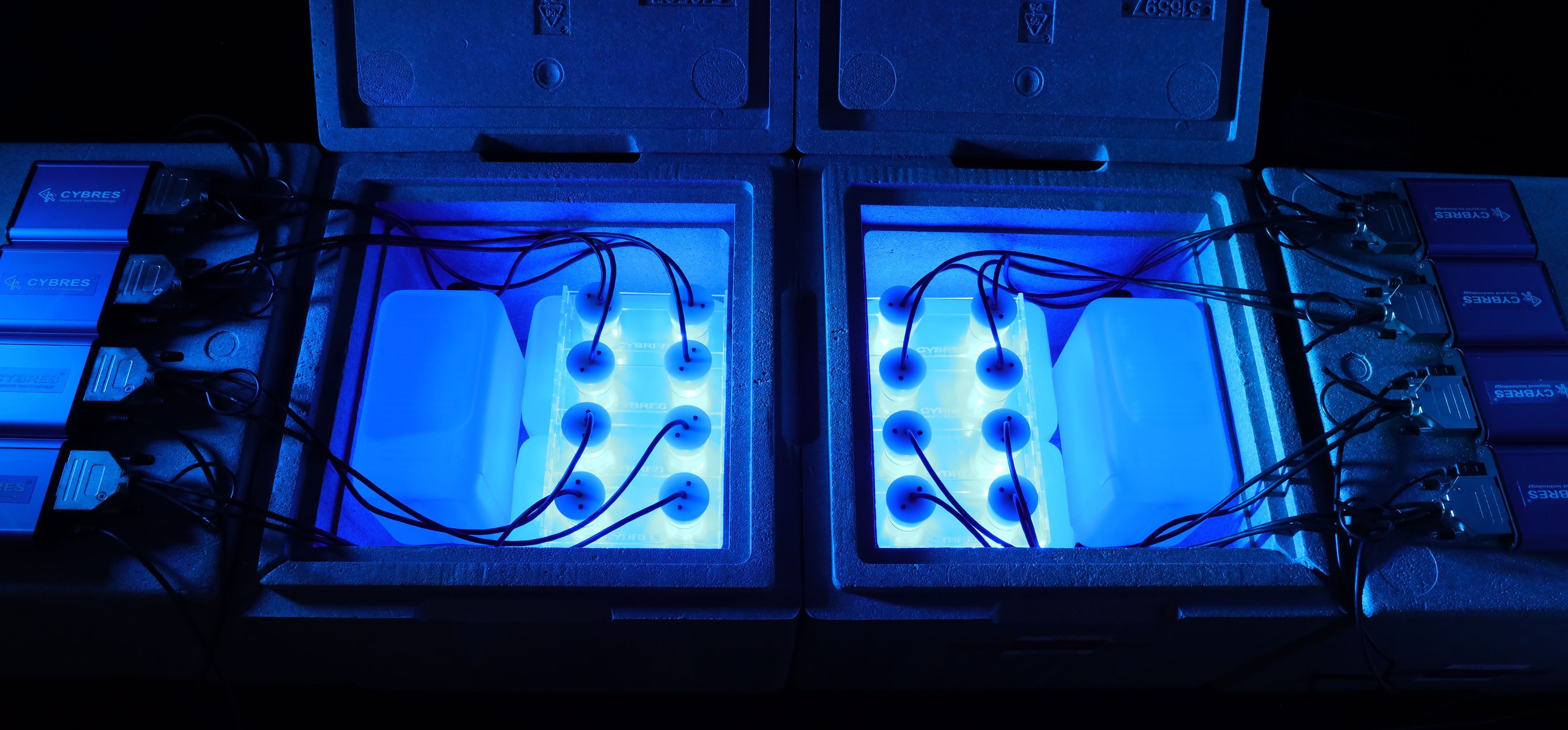}}
\caption{\textbf{(a)} Four EIS cells with 15 ml of pure water, this setup is installed in a thermo-insulating container. \textbf{(b)} Electrochemical cell with openings for \ce{CO_2} input; \textbf{(c)} Setup with 16 electrochemical cells within \ce{CO_2} scenarios placed in two thermo-insulating containers, the distance between closest rows of cells in different containers is about 150mm. Each thermo-insulating container has 4.5 liters of water inside for thermal stabilization. Illumination of cells by 490nm light is on for demonstration purposes (in experiments the light is off). \label{fig:CO2_scenario}}
\end{figure*}

Coupling between such chaotic electrochemical oscillators is of further interest. Here we need to distinguish two different scenario: synchronization of electrochemical oscillators at short distances inside one fluidic cell, and between different cells. In the first case we observe large-amplitude EIS waves from a single electrochemical cell, the second case is characterized by a synchronization of such waves between several cells. Short-distance scenario includes molecular coupling mechanisms, among them exchange of protons in hydrogen-bound networks \cite{Konyukhov11} leading to flip-flop processes of spin states. Physical implementation of cell-cell couplings is discussed in several works \cite{Frimmer17}, which propose light-matter interactions, Rabi oscillations (dynamics of a spin in a magnetic field), spin-based interactions \cite{Trukhanova23} or photon-assisted entanglement \cite{Zhang20}. 

These considerations can be exemplified by electrochemical dynamics shown in Fig. \ref{fig:singleWave}. Typically, short-term chaotic fluctuations are uncorrelated and correspond to the chaotic oscillation mode in Fig. \ref{fig:oscillatorScheme}. Under certain conditions, individual electrochemical cells produce large-amplitude waves -- this reflects a synchronous dynamics of reactions (\ref{eq:carbonicAcid})-(\ref{eq:carbonicAcidIonsFurther}) inside a single cell. These waves can be also correlated with each other, here we observe a synchronization of molecular oscillators between cells. Considering environmental parameters, see Fig. \ref{fig:shortTermCorrAdd}, we do not observe their correlations with  electrochemical waves.

In experiments with multiple fluidic cells, individual large-amplitude waves from separate cells are not always correlated with all other cells. We discuss this case in Sec. \ref{sec:severalContainers}. Since para-/ortho- conversion affects not only electrochemical reactivity, but also a heat capacity  \cite{kernbach23Thermal}, we expect the appearance of temperature-impedance waves in fluidic cells. This effect is also observed in experiments and is discussed in Sec. \ref{sec:TempImpWaves}.

\section{Setup, methodology and analysis}

\subsection{Setup}

Experiments are performed in groups of four, six, eight and sixteen fluidic cells placed in holders at edge-to-edge distance of 60mm (diameter of cells 26.5 mm, material -- PE Polypropylen), see Fig. \ref{fig:experimentalSetup}. Each cell has one electrochemical and one temperature sensors immersed into about 12ml (within \ce{CO_2} scenario) or 15 ml (closed cells) of distilled water with initial conductivity 0.02 $\mu S/cm$. Electrochemical cells are processed pairwise by one EIS device, all setups use independent computational units and independent power supplies. Each computational unit, in addition to EIS and  temperature of fluids, collects data from 3D accelerometer/magnetometer, sensor of EM emission in 450MHz-2.5GHz range, air pressure sensor, environmental humidity/temperature and \ce{CO_2} sensor -- totally 26 sensor data channels with time stamp, synchronized between different EIS devices. Electrochemical cells for \ce{CO_2} scenario have two openings, see Fig. \ref{fig:CO2_scenario_A}, and follow the \ce{CO_2}-controlled/uncontrolled methodology used in \cite{kernbach2022electrochemical}. The \ce{CO_2} level is continuously measured and is typically at 1400-1000 ppm at the begin of measurements. All experiments are grouped into three series, see Table \ref{tab:parametersSetup}, reflecting combinations of \ce{CO_2} access and transparent/non-transparent setups. To test synchronization of molecular oscillators by \ce{CO_2} level, electrochemical cells are installed in two or three different non-transparent thermo-insulating containers, see Fig. \ref{fig:CO2_scenario_B}.

\begin{table}[ht]
\centering
\caption{\small Parameters of setup, EIS $f_{excitation}=450Hz$, passive thermostabilization, all additional sensors are on. \label{tab:parametersSetup}}
\fontsize {9} {10} \selectfont
\begin{tabular}{
p{0.8cm}@{\extracolsep{3mm}}
p{3.2cm}@{\extracolsep{3mm}}
p{2.0cm}@{\extracolsep{3mm}}
p{1.5cm}@{\extracolsep{3mm}}
}\hline
series& description                    & N of containers  & N of cells \\\hline
1     & no \ce{CO_2}  								 & 1,2 & 4-12 \\
2     & non-transparent, \ce{CO_2}     & 1,2 & 4,6  \\
3     & transparent, \ce{CO_2}         & 1-3 & 4-16 \\
\hline
\end{tabular}
\end{table}

\begin{figure}[htp]
\centering
\includegraphics[width=0.49\textwidth]{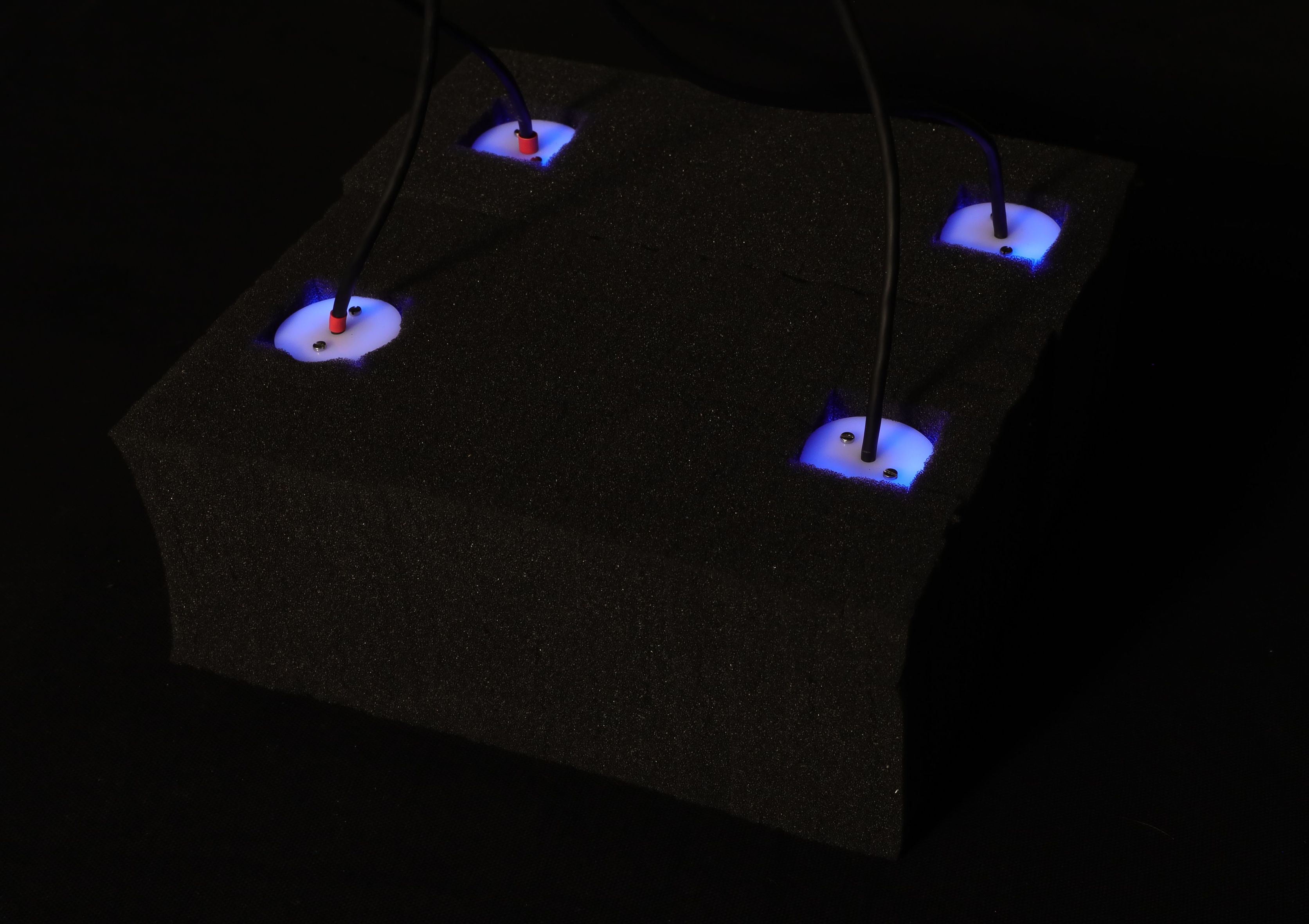}
\caption{Optically non-transparent setup with four cells inserted into the foam. \label{fig:lightIsolatedSetup}}
\end{figure}

In parallel to optically-transparent setup shown in Fig. \ref{fig:experimentalSetup}, we tested optically non-transparent setups, as shown in Fig. \ref{fig:lightIsolatedSetup}. Here fluidic cells are inserted into the foam that enables the \ce{CO_2} access; all cells are placed into one thermo-insulated container. Comparison between one and several non-transparent setups should clarify how the optical transparency and common \ce{CO_2} level affect achieving the synchronization events.    

\subsection{Methodology and analysis}

Correlations are calculated for impedance and temperature data obtained from EIS devices in real time. To remove trends, the original data $data(x)$ are approximated by a nonlinear function of $n$-order with coefficients $k$
\begin{equation}
\label{eq:approx2}
fit(x)=k_nx^n+n_{n-1}x^{n-1}+...+k_3x^3+k_2x^2+k_1x+k_0
\end{equation}
using the Levenberg-Marquardt algorithm \cite{Dennis83}, where we calculate the residual curve
\begin{equation}
\label{eq:resifual}
m=data(x)-fit(x).
\end{equation}
Considering $m^{(1)}_i$ and $m^{(2)}_i$ as $i$-time step samples from two impedance-impedance, impedance-temperature or temperature-temperature channels, the transformation (\ref{eq:resifual}) ensures a linear dependency between  $m^{(1)}_i$ and $m^{(2)}_i$ in the same scale (due to removal of different trends) that is useful for determining the Pearson's linear correlation coefficient. For calculation we follow the algorithm from \cite{Press:1992:NRC:148286}
\begin{equation}
\label{eq:correlation}
r^{m^{(1)},m^{(2)}}=\frac{\sum_i{(m^{(1)}_i-\bar{m}^{(1)})(m^{(2)}_i-\bar{m}^{(2)})}}{\sqrt{\sum_i(m^{(1)}_i-\bar{m}^{(1)})^2}\sqrt{\sum_i(m^{(2)}_i-\bar{m}^{(2)})^2}},
\end{equation}
where $\bar{m}^{(1)}$ is the mean of the $m^{(1)}_i$'s, $\bar{m}^{(2)}$ is the mean of the $m^{(2)}_i$'s. All $r^{m^{(1)},m^{(2)}}$ are calculated as rolling correlations within the window of size $t_{synch}$. 

\begin{figure}[htp]
\centering
\subfigure{\includegraphics[width=0.49\textwidth]{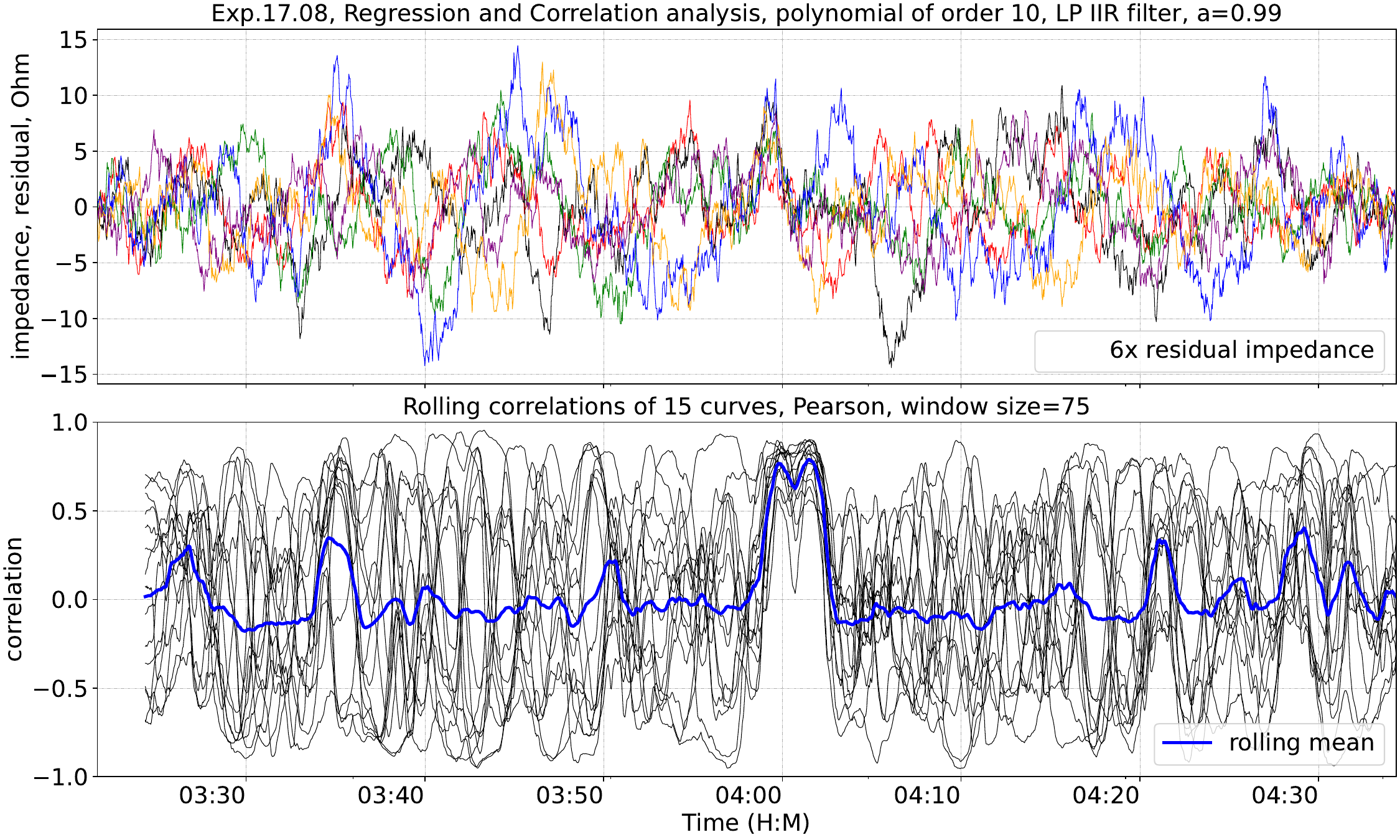}}
\caption{Exp.17.08, 6 electrochemical oscillators in one thermostabilizing container in \ce{CO_2} scenario; \textbf{(upper)} 6x residual impedance curves; \textbf{(lower)} 15x rolling correlations, blue curve -- mean of all rolling correlations (\ref{eq:rollMean}) with max. $r=0.79$, see Fig.\ref{fig:size_vs_corr}. Polynomial (\ref{eq:approx2}) of 10$^{th}$ order is used. \label{fig:exp1708}}
\end{figure}

\begin{figure}[htp]
\centering
\subfigure{\includegraphics[width=0.49\textwidth]{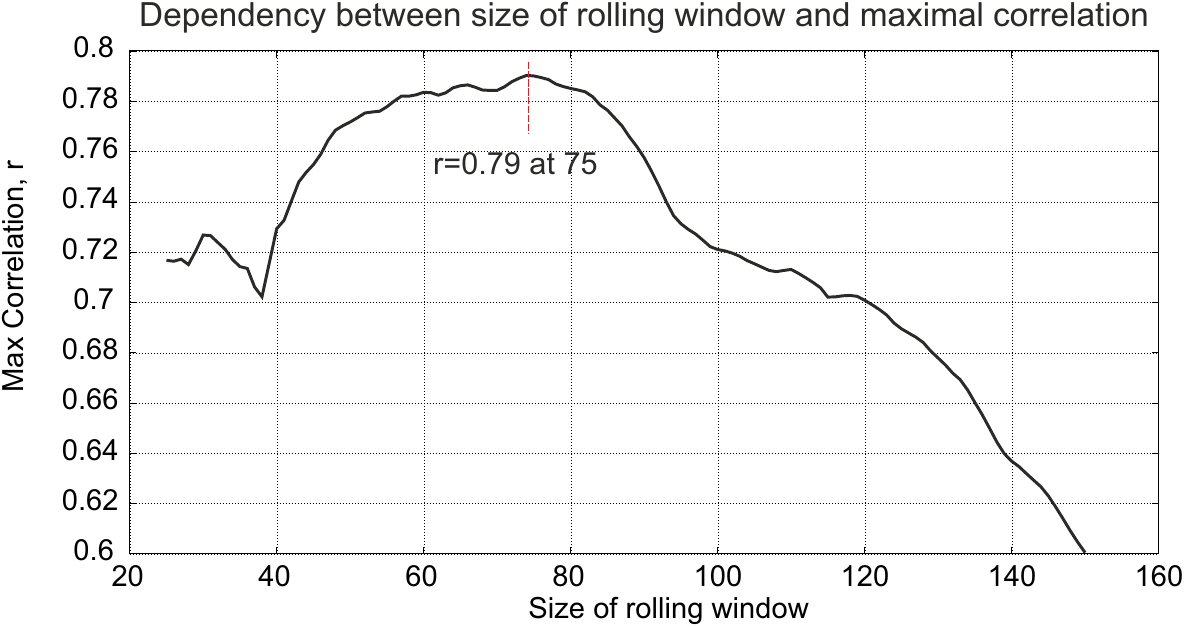}}
\caption{Dependency between size of rolling window $t_{synch}$ and maximal correlation $r^{mean}_i$ from Fig. \ref{fig:exp1708}. Polynomial (\ref{eq:approx2}) of 10$^{th}$ order is used.\label{fig:size_vs_corr}}
\end{figure}

For calculation of correlations between multiple independent variables we use ideas of 'a multi-way correlation coefficient' \cite{taylor2020multi} based on eigenvalues of symmetrical correlation matrices
\begin{equation}
\label{eq:correlationMatrix}
\begin{bNiceMatrix}[light-syntax,first-row,first-col]
{} a b c d;
a 1  r_{a,b}  r_{a,c} r_{a,d};
b .  1 r_{b,c} r_{b,d};
c . . 1 r_{c,d};
d . . .  1
\end{bNiceMatrix}
\end{equation}
where $a, b, c, d$ are independent variables (the coefficient $R^2$ of multiple correlation used for estimating predictability of the dependent variable from independent variables cannot be applied here). If $p$ is the number of independent oscillators, it needs to analyze $C(p,2)$ correlation curves -- combination of 2 from $p$ (pairwise correlations from $p$ independent electrochemical oscillators), e.g. $C(6,2)=15$, $C(8,2)=28$, $C(12,2)=66$, etc. Due to large number of rolling correlations, it makes sense to calculate their rolling mean:
\begin{equation}
\label{eq:rollMean}
r^{mean}_i=\frac{1}{C(p,2)} \sum^n{r_i^{v,g}}~ |~ (v,g)~\in ~C(p,2)
\end{equation}
Synchronization process manifests when all or several pairs $r^{m^{(1)},m^{(2)}}$ become correlated, this is observable as a peak of rolling mean, see Fig. \ref{fig:exp1708}. Exact values of $r^{mean}_i$ depends on the selected size $t_{synch}$ of rolling window, as shown in Fig. \ref{fig:size_vs_corr}. Typically $t_{synch}$ lies between 70 and 200 samples. To find a maximal value of rolling mean, the program additionally scans $t_{synch}$.

\begin{figure}[htp]
\centering
\subfigure{\includegraphics[width=0.49\textwidth]{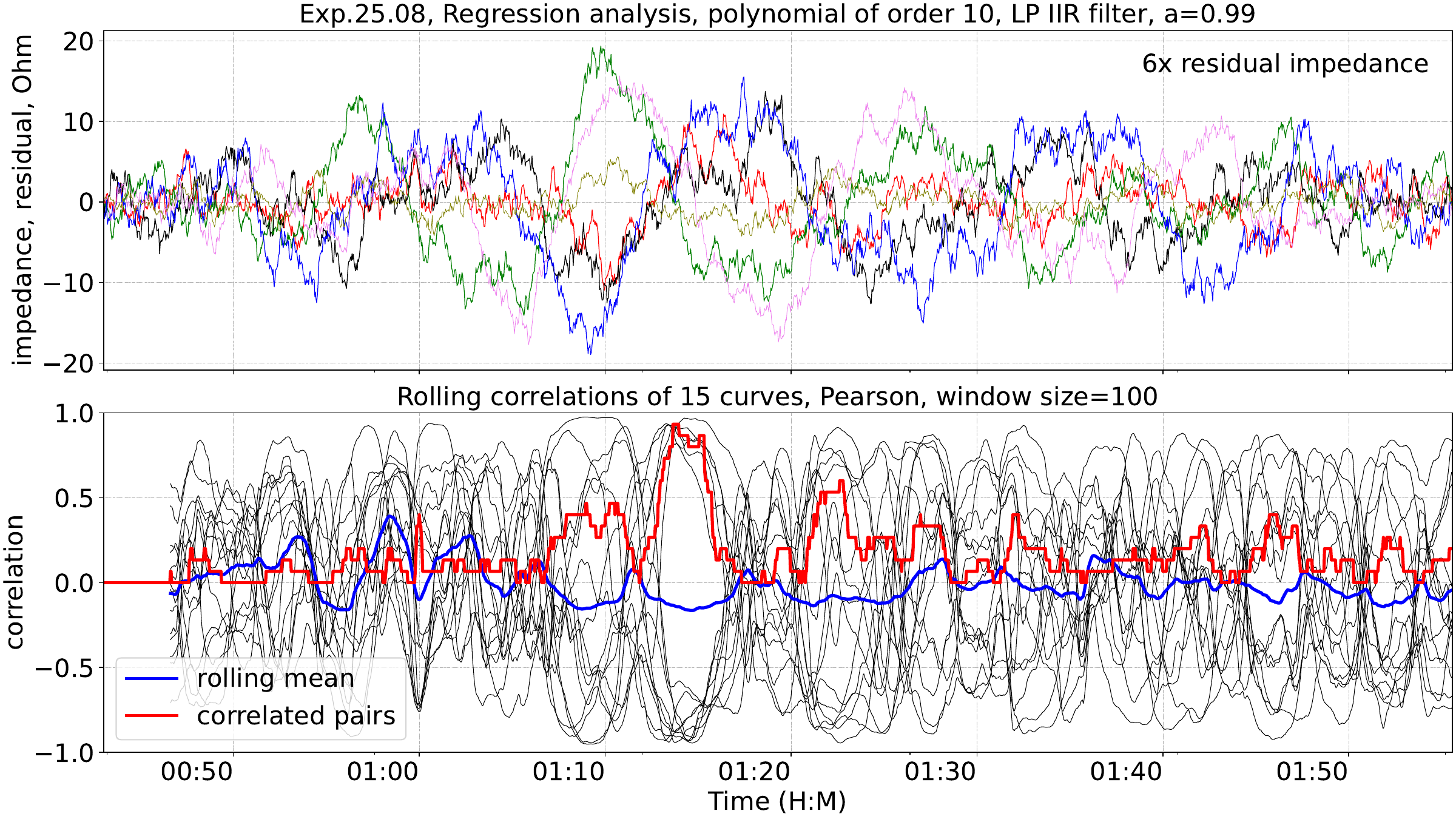}}
\caption{Exp.25.08, comparison of metrics with $l^{pair}_i$ and $r^{mean}_i$ in (\ref{eq:finalMetrix1}); \textbf{(upper)} 6x residual impedance curves; \textbf{(lower)} 15x rolling correlations, blue curve -- mean of all rolling correlations (\ref{eq:rollMean}), red curve --   correlated pairs of electrochemical oscillators (\ref{eq:finalResult}) with max. $l_i^{pair}=0.93$. \label{fig:correlated_pairs}}
\end{figure}

As already mentioned, cells are not always fully correlated with each other or are correlated in anti-phase; in such cases $r^{mean}$ does not provide a useful information about overall dynamics. Here the synchronization of independent chaotic oscillations can be expressed in degree of freedoms -- $l$ pairs of electrochemical oscillators $m^{(1)}$ and $m^{(2)}$ (related to all pairs $C(p,2)$) that simultaneously demonstrate high values of $r^{m^{(1)},m^{(2)}}$. It can be formalized as the following algorithm:
\begin{equation}
\label{eq:finalResult}
l^{pair}_i: ~~ \frac{1}{C(p,2)} \cdot \forall \sum^n{|r_i^{v,g}|}>z~~ |~ (v,g)~\in ~C(p,2)
\end{equation}
where $z$ is a correlation threshold (e.g. $z=0.7$). Synchronization event is detected if 
\begin{equation}
\label{eq:finalMetrix1}
r^{mean}_i\geq z ~~~~~~ or ~~~~~~~~~ l^{pair}_i \approx 1
\end{equation}
Expression (\ref{eq:finalMetrix1}) represents two metrics, which reflect different properties of overall dynamics, e.g. $l^{pair}_i$ is useful in cases of anti-correlations; closeness to 1 (e.g. $l^{pair}_i>0.9$, $l^{pair}_i>0.8$, $l^{pair}_i>0.7$, see Table \ref{tab:controlMeasurements}) demonstrates the degree of pairwise mutual synchronization, see Fig \ref{fig:correlated_pairs}. 

Since all attempts have different durations $t_{exp}$, depending on electrochemical degradation of samples, their numerical evaluation has been conducted as the number of samples $N_{smp}$ satisfying the metrics (\ref{eq:finalMetrix1}) divided by $t_{exp}$
\begin{equation}
\label{eq:finalMetrix2}
\frac{1}{t_{exp}} N_{smp}~\forall~(r^{mean}_i\geq z) ~~~~~~ or ~~~~~~~ \frac{1}{t_{exp}} N~ \forall~(l^{pair}_i \approx 1)
\end{equation}
This metric is shown in Table \ref{tab:controlMeasurements}. 

Evaluation metric based on (\ref{eq:finalResult}) has a specific dependency between the number $N_{pairs}$ of $v,g$ pairs satisfying $r_i^{v,g}>z$ or $|r_i^{v,g}|>z$ and the number of cell $N_{cells}$ involved into these pairs. For instance three pairs 1-2, 2-3, 3-4 involve four cells 1, 2, 3, 4. The ratio 
\begin{equation}
\label{eq:finalMetrix3}
\frac{N_{pairs}}{N_{cells}} \leq \frac{C(N_{cells},2)}{N_{cells}}
\end{equation}
is characterized by $\{1.5, 2.5, 3.5, 4.5,...\}$ for $N_{cells}=\{4, 6, 8, 10,...\}$ and is representative for evaluating the degree of mutual synchronization between $N_{cells}$ cells, see Fig. \ref{fig:subSetAnalysis}. 

We implemented two versions of algorithms for (\ref{eq:approx2})--(\ref{eq:finalMetrix3}) to avoid computational artifacts: the first one is implemented manually in C++, the second one uses high-efficient Python libraries like NumPy. For selecting the $n$-order function (\ref{eq:approx2}) for regression it is important to achieve the best possible fit. Typically for the data frame of 2-4 hours (1500-5000 samples for a single sampling within 1-5 sec.) we used the polynomial (\ref{eq:approx2}) of 5$^{th}$ order in initial experiments, which produces well observable synchronization waves, see Fig. \ref{fig:singleWave}. However, due to better linearity of data for the correlation analysis, $n=10$ for (\ref{eq:approx2}) is set for all later experiments, which produces flat residual dynamics. The regression analysis increases a dynamic range, the appeared noise should be limited by low-pass filter. The implemented IIR filters do not essentially impact the calculation of $r$ as long as the cut-off frequency is not too low (the signal waveform is not distorted).

\section{Control measurements}

\subsection{Electrochemical dynamics without \ce{CO_2} access}

Control measurements without \ce{CO_2} access have been conducted in setups shown in Figs. \ref{fig:CO2_scenario}. 

\begin{figure}[htp]
\centering
\includegraphics[width=0.49\textwidth]{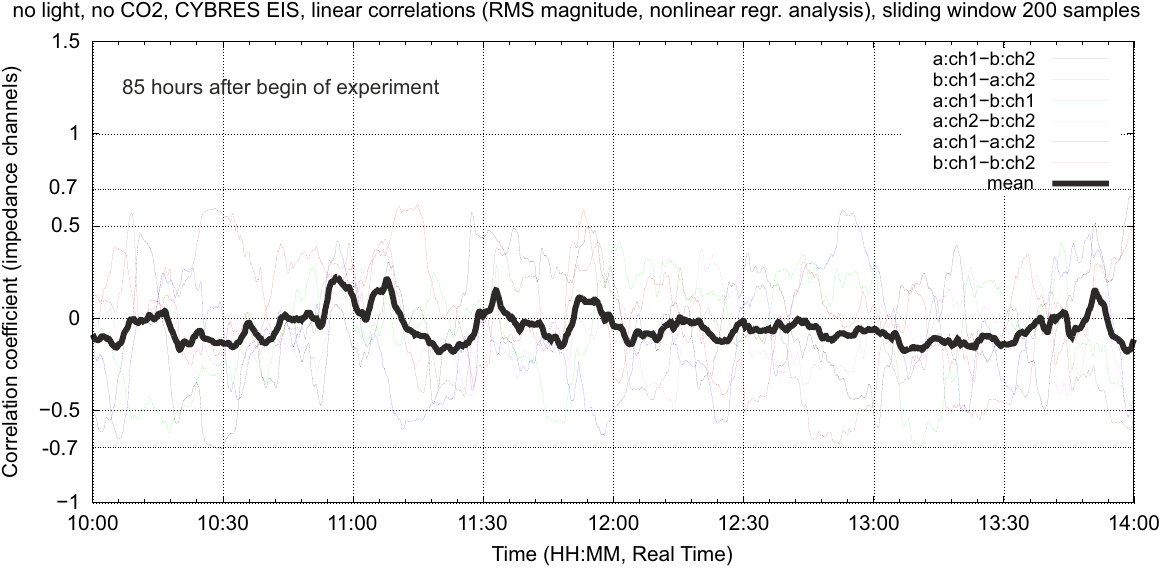}
\caption{Example of control measurements without \ce{CO_2} access during 4 hours, shown are six $r^{m^{(1)},m^{(2)}}$ from four EIS channels with rolling mean $r^{mean}_i$. \label{fig:control}}
\end{figure}

Example of such measurements is shown in Fig. \ref{fig:control} -- typically, $r^{m^{(1)},m^{(2)}}$ have a low amplitude, multiple correlation curves fulling the conditions (\ref{eq:finalMetrix1}) are not observed. Since water samples without \ce{CO_2} demonstrate slow electrochemical degradation, control experiments and their statistical evaluation have been performed in long-term (several weeks) and short-term (several days, similar to experimental runs) attempts, and accumulated in Table \ref{tab:controlMeasurements}. Synchronization effects are not observed in 8 and more cells, 6 cells demonstrate rare synchronization events.

\begin{table*}[ht]
\centering
\caption{\small Overview of control attempts (without \ce{CO_2}) and comparison with experimental results\\ (with \ce{CO_2}) of equivalent size, data frame 2000 samples, running window 100, shown is the metric (\ref{eq:finalMetrix2}). \label{tab:controlMeasurements}}
\fontsize {9} {10} \selectfont
\begin{tabular}{
p{0.5cm}@{\extracolsep{3mm}}
p{1.8cm}@{\extracolsep{3mm}}
p{1.5cm}@{\extracolsep{3mm}}
p{1.6cm}@{\extracolsep{3mm}}
p{1.6cm}@{\extracolsep{3mm}}
p{1.6cm}@{\extracolsep{3mm}}
p{1.6cm}@{\extracolsep{3mm}}
}\hline
N cells & experiment duration, sec & N of samples, $\times 10^6$ & \multicolumn{4}{c}{{evaluation metric}} \\\hline
        &                           &                   & $r^{mean}_i>$0.7 & $l^{pair}_i>$0.9 & $l^{pair}_i>$0.8 & $l^{pair}_i>$0.7\\\hline \hline \\[-2mm]
\multicolumn{7}{c}{\textit{Long-term control attempts}}\\[2mm]
12 & 1427624   & 7.989        & 0           & 0                    & 0                    & 0                \\
10 & 1427700   & 6.658        & 0           & 0                    & 0                    & 9.105e-6         \\
8  & 1427707   & 5.326        & 0           & 0                    & 3.502e-6             & 2.661e-5         \\
6  & 1431234   & 4.133        & 6.288e-5    & 1.676e-5             & 3.772e-5             & 2.620e-4         \\
6  & 1431847   & 4.131        & 5.796e-5    & 4.330e-5             & 9.707e-5             & 3.114e-4         \\
6  & 1427707   & 3.994        & 5.883e-5    & 5.813e-5             & 1.078e-4             & 4.363e-4         \\\hline\\[-2mm]
\multicolumn{7}{c}{\textit{Short-term control attempts with 8 cells} }\\[2mm]
8  & 287916   & 1.040         & 0           & 0                    & 0                    & 3.820e-5         \\
8  & 441099   & 1.743         & 0           & 0                    & 1.133e-5             & 3.627e-5         \\
8  & 441142   & 1.740         & 0           & 0                    & 0                    & 2.493e-5         \\\hline
 \multicolumn{3}{r}{\textit{mean}}     & \textbf{0}  & \textbf{0}           & \textbf{3.777e-6}  & \textbf{3.313e-5}  \\\hline \\[-2mm]
\multicolumn{7}{c}{\textit{Experiments with 8 cells}}\\[2mm]
8  & 231395    & 0.939        & 9.507e-5    & 0                    & 0                    & 1.253e-4\\
8  & 232427    & 0.941        & 1.075e-4    & 4.732e-5             & 1.850e-4             & 7.012e-4\\
8  & 658559    & 2.503        & 6.073e-6    & 0             			 & 9.110e-6             & 8.047e-5\\\hline
 \multicolumn{3}{r}{\textit{mean}}     & \textbf{4.949e-5}  & \textbf{1.577e-5} & \textbf{6.470e-5} & \textbf{3.023e-4} \\\hline \\[-2mm]
\multicolumn{7}{c}{\textit{Short-term control attempts with 6 cells}}\\[2mm]
6  & 257370   & 0.870         & 5.439e-5    & 0                    & 7.770e-6             & 6.216e-5         \\
6  & 266101   & 0.910         & 0           & 9.019e-5             & 1.653e-4             & 5.599e-4         \\
6  & 441099   & 1.401         & 4.534e-5    & 4.760e-5             & 9.068e-5             & 3.083e-4         \\
6  & 441142   & 1.402         & 1.450e-4    & 7.707e-5             & 1.292e-4             & 6.256e-4         \\
6  & 426685   & 1.391         & 0           & 8.905e-5             & 1.828e-4             & 5.554e-4         \\
6  & 426826   & 1.391         & 0           & 1.640e-5             & 3.748e-5             & 1.780e-4         \\\hline
 \multicolumn{3}{r}{\textit{mean}}     & \textbf{4.079e-5}  & \textbf{5.339e-5} & \textbf{1.022e-4}  & \textbf{3.816e-4} \\ \hline \\[-2mm]
\multicolumn{7}{c}{\textit{Experiments with 6 cells}}\\[2mm]
6  & 231395    & 0.939        & 7.346e-5    & 0                    & 3.025e-5             & 2.895e-4\\
6  & 658559    & 1.877        & 1.943e-4    & 2.201e-4             & 3.780e-4             & 1.065e-3\\
6  & 462169    & 1.339        & 3.873e-4    & 1.428e-4             & 2.899e-4             & 1.001e-3\\
6  & 232297    & 0.939        & 2.884e-4    & 5.381e-4             & 8.437e-4             & 2.410e-4\\
6  & 462169    & 1.339        & 2.336e-4    & 1.384e-4             & 3.310e-4             & 8.157e-4\\
6  & 462169    & 1.339        & 2.077e-4    & 1.579e-4             & 2.769e-4             & 7.810e-4\\\hline
\multicolumn{3}{r}{\textit{mean}}      & \textbf{2.0308e-4} & \textbf{1.996e-4} & \textbf{3.583e-4}  & \textbf{6.989e-4}\\
\hline
\end{tabular}
\end{table*}

\subsection{Correlations between impedance and temperature of fluids}
\label{sec:controlTemperature}

\begin{figure}[htp]
\centering
\subfigure{\includegraphics[width=0.49\textwidth]{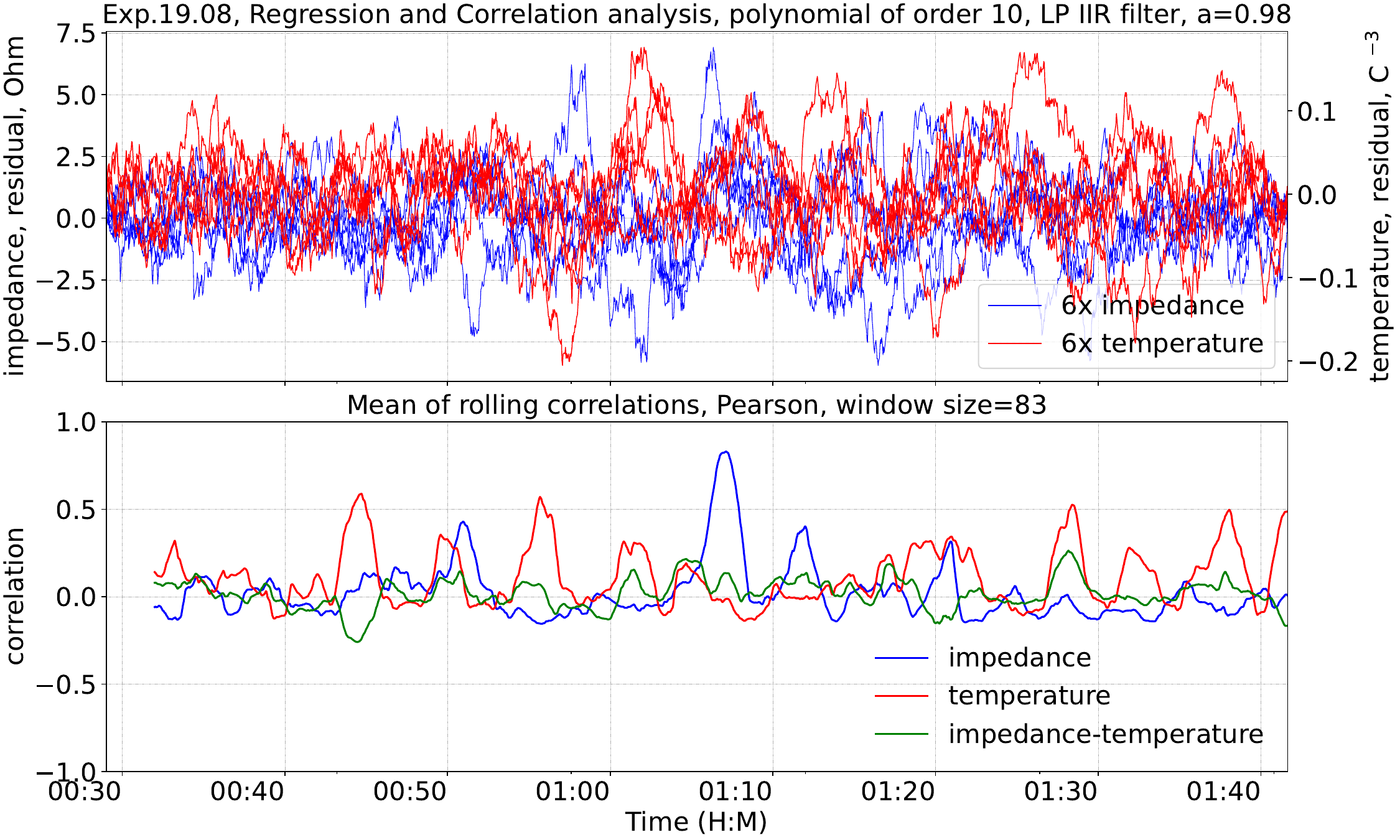}}
\caption{Exp.19.08 with six 6 electrochemical oscillators in one thermostabilizing container in \ce{CO_2} scenario; \textbf{(upper)} Residual impedance and temperature dynamics -- 6 impedance curves (blue) and 6 fluidic temperature curves (red);  \textbf{(lower)} Means of rolling correlations in three groups: 15 impedance-impedance correlations (blue), 15 temperature-temperature correlations (red), and 36 impedance-temperature correlations (green). Correlations of temperature and impedance do not overlap. \label{fig:corr_temp_imp2}}
\end{figure}

\begin{figure}[htp]
\centering
\subfigure{\includegraphics[width=0.49\textwidth]{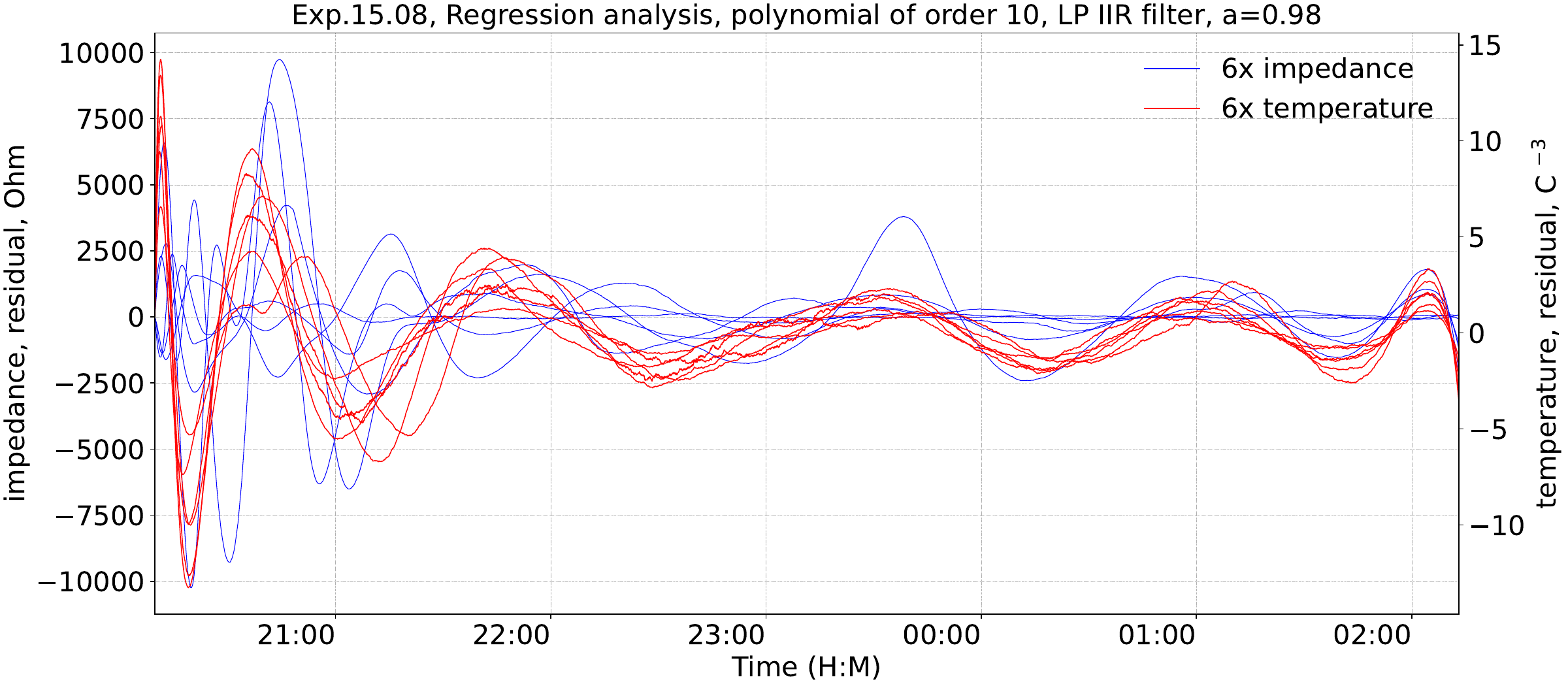}}
\caption{Oscillating electrochemical and thermal dynamics of \ce{CO_2} dissolving in initial stages of experiments. \label{fig:initialOscillations}}
\end{figure}

Daily temperature dynamics in laboratory penetrates into the containers and represents a common factor influencing all cells. Thus, the temperature can be responsible for the observed synchronization effects. However, due to thermo-insulating containers, such correlated temperature variations (deviations from trend) inside fluidic cells are very small and slow -- on the level of $10^{-3}-10^{-4}~ ^\circ C$ per several hours, see Figs. \ref{fig:corr_temp_imp2}, \ref{fig:initialOscillations}. Following known dependencies between temperature and electrical conductivity \cite{Hayashi2004}
\begin{equation}
\label{eq:t}
EC_t=EC_{25}[1+a(t_{25})],
\end{equation}
where $a$ varies between 0.0191 and 0.025, $EC_t$ is a conductivity at temperature $t$, $EC_{25}$ is a conductivity at 25$^\circ C$, we expect such small temperature-driven variations about $10^{-5}-10^{-6}$ of conductivity close to 25$^\circ C$; in term of impedances, the slow temperature dynamics can generate about 0.1-1 Ohm (for 100 kOhm at 25$^\circ C$). However, we observe variations about 10-30 Ohm; thus, the daily temperature rhythms cause impedance changes about one-two orders of magnitude smaller than observed in experiments. Fig. \ref{fig:exp050222Add} exemplifies this consideration for the case of two cells.

To test the dependency between impedance and temperature, we use impedance and temperature channels of fluidic cells. Rolling correlations of all residual curves are calculated in three groups -- as impedance-impedance, temperature-temperature, and impedance-temperature correlations. Overlapping between these groups indicates a temperature-driven synchronization. Fig. \ref{fig:corr_temp_imp2} demonstrates residual dynamics of impedance and temperature of six cells (6 impedance curves and 6 fluidic temperature curves) with rolling mean of 15 impedance-impedance, 15 temperature-temperature, and 36 impedance-temperature correlations. Temperature and impedance correlations do not overlap, suggesting synchronization effects that are not due to temperature. In further analysis we always calculate the rolling means of impedance-temperature correlations to test and to avoid temperature-related artifacts. There are two cases when the temperature can affect the EIS dynamics: fast changes of temperature (they affect control measurements in Table \ref{tab:controlMeasurements} among other environmental parameters) and appearance of temperature-impedance waves, discussed in Sec. \ref{sec:TempImpWaves}.

\subsection{Oscillating reactions in \ce{CO_2} dissolving}

Different oscillating parameters of water have been found by optical, electrochemical or NMR spectroscopy; some of them are assumed to be related to spin isomers \cite{Morre16, Drozdov14}. Other periodical oscillations are related to daily temperature cycles. Dissolving of \ce{CO_2} can also generate oscillating electrochemical and thermal dynamics, especially in initial stages of experiments with duration up to 8-12 hours, see Fig. \ref{fig:initialOscillations}.

\begin{figure}[htp]
\centering
\subfigure[]{\includegraphics[width=0.49\textwidth]{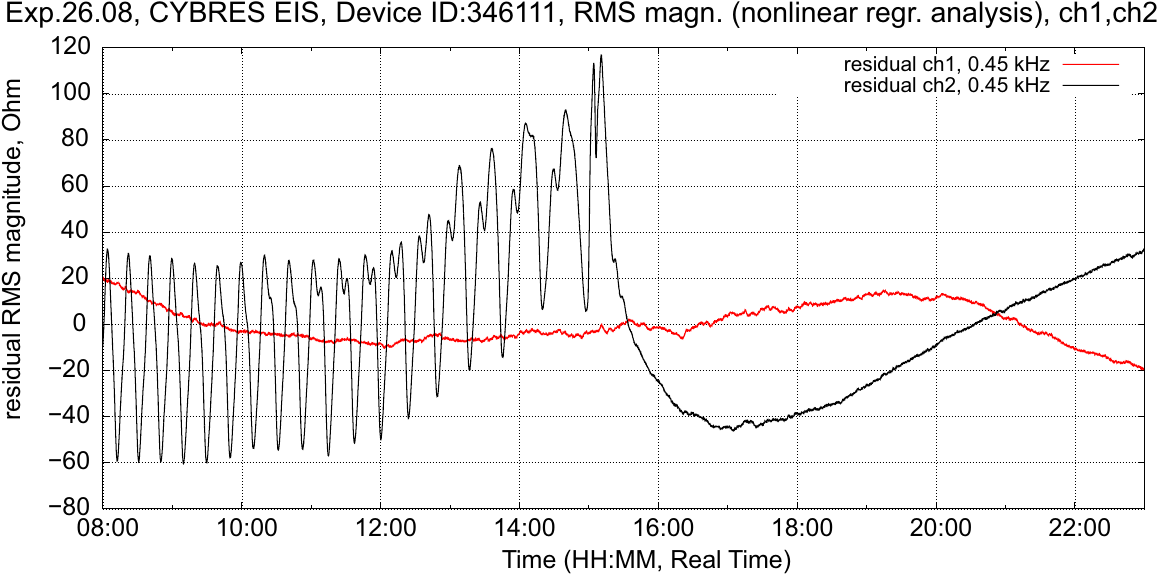}}
\caption{Oscillating electrochemical dynamics within \ce{CO_2} dissolving scenario, shown in the stage of spontaneous stop of oscillations. The channel 1 (red curve) does not demonstrates any oscillating dynamics. \label{fig:initialOscillations6}}
\end{figure}

These oscillations possess decreasing amplitudes and are correlated with each other due to a common triggering event. To avoid detection of such artifacts, the analysis does not consider first 18-24 hours of measurements. In addition to initial stages of experiments, self-oscillatory effects can occur randomly on any EIS channel and are characterized by a stable period and large amplitude with a spontaneous start and stop of oscillations, see Fig. \ref{fig:initialOscillations6}. This self-oscillating behavior is of further interest since it can be also triggered by cross-photon excitation and spin conversion mechanism. However, such an oscillating dynamics is not included in analysis because it is not related to reactions (\ref{eq:carbonicAcid})-(\ref{eq:carbonicAcidIonsFurther}) and reflects different generating pathways. Analysis software automatically rejects large-amplitude oscillations.

\section{Experimental results}

\begin{figure}[htp]
\centering
\subfigure{\includegraphics[width=0.49\textwidth]{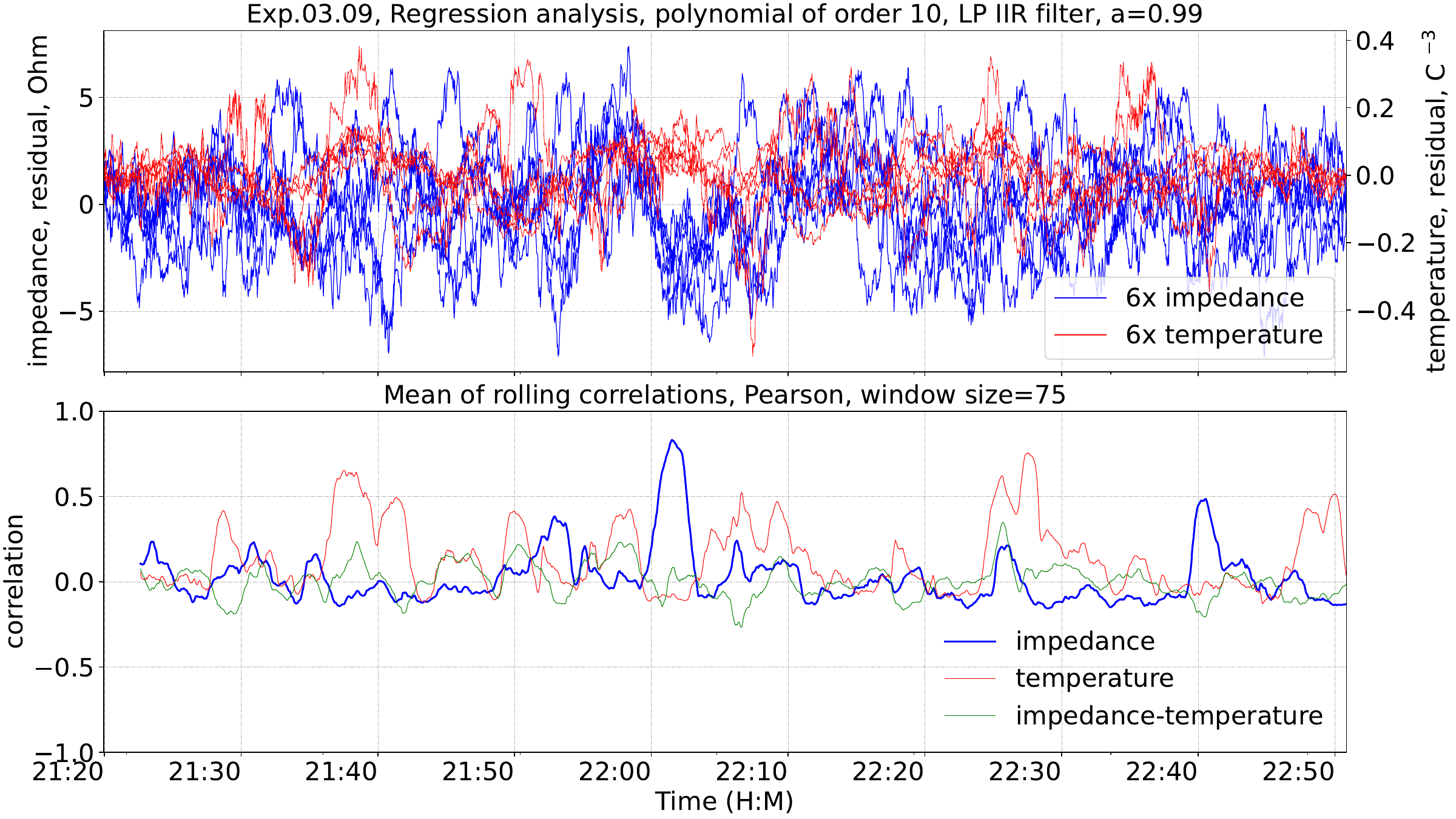}}
\caption{Example of a synchronization event between six cells placed in one thermo-insulating container; \textbf{(upper)} Residual impedance and temperature dynamics of six cells; \textbf{(lower)} Means of rolling correlations in three groups, maximum of impedance (blue curve) $r_{mean}=0.84$. \label{fig:exp0309_temp_imp}}
\end{figure}

Typical synchronization events are shown in Figs. \ref{fig:corr_temp_imp2}, \ref{fig:exp0309_temp_imp}, we observe several individual waves (from each cell) that have
the same phase and thus produce a peak of mean rolling correlation in impedance-impedance group. At the same time, temperature curves do not demonstrate any correlated dynamics, i.e the influence of temperature in such events can be excluded.

Statistical evaluation has been conducted with 6 and 8 cells with $>10^7$ samples, several control and experimental attempts with equal sample size are collected in Table \ref{tab:controlMeasurements}. We observe about order of magnitude difference between them based on $r^{mean}_i>0.7$ and $l^{pair}_i>0.9$ metrics for 6 cells and a qualitative difference for 8 cells.

\subsection{High correlation in 4 and 6 cells}
\label{sec:highCorrealtions}

\begin{figure}[htp]
\centering
\subfigure{\includegraphics[width=0.49\textwidth]{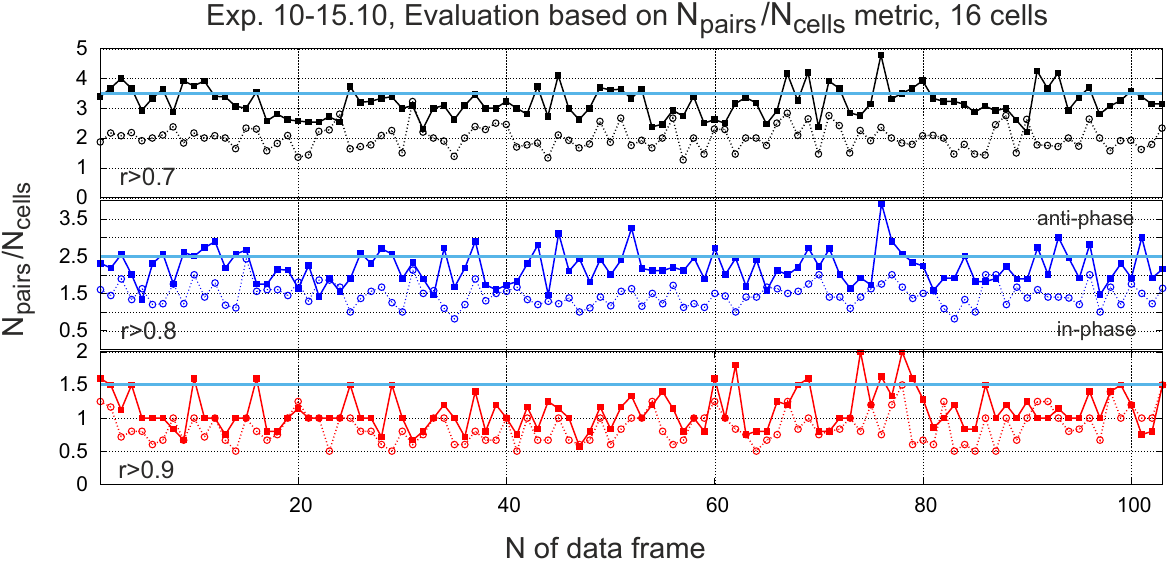}}
\caption{Evaluation based on $N_{pairs}/N_{cells}$ metric, data frame 1500 samples, 16 cells, solid line -- $|r_i^{v,g}|>z$ (anti-phase synchronized oscillations), dashed line -- $r_i^{v,g}>z$ (in-phase synchronized oscillations), shown are 1.5, 2.5 and 3.5 characteristic values. \label{fig:subSetAnalysis}}
\end{figure}

Setups with a large number of cells have subsets of 4 and 6 cells with a high correlation (max. $r^{mean}_i>0.9$) between them. Dynamics of such subsets is analyzed for $r_i^{v,g}>z$ (in-phase synchronized oscillations, see Fig. \ref{fig:exp1708}) and $|r_i^{v,g}|>z$ (anti-phase synchronized oscillations, see Fig.\ref{fig:correlated_pairs}) in (\ref{eq:finalResult}) for three cases of $r>0.9$, $r>0.8$ and $r>0.7$, see Fig. \ref{fig:subSetAnalysis}. Evaluation metric is based on (\ref{eq:finalMetrix3}) with $\frac{N_{pairs}}{N_{cells}}$ ratio. 

We see that anti-phase synchronized oscillations are more frequent and have higher correlations. The case with $r>0.9$ is primarily related to synchronization events between four cells that occur approximately once every 8000 samples, $r>0.8$ and $r>0.7$ -- to 6, 8 and even 10 cells with a synchronization event every 3000-3500 samples. In general, we do not see a clear pattern of appearance (e.g. at the beginning of experiment), synchronization events tend to have higher and lower appearance periods randomly distributed over the course of experiment.

\subsection{Desynchronization in non-transparent setups}
\label{sec:severalContainers}

\begin{figure}[htp]
\centering
\subfigure{\includegraphics[width=0.49\textwidth]{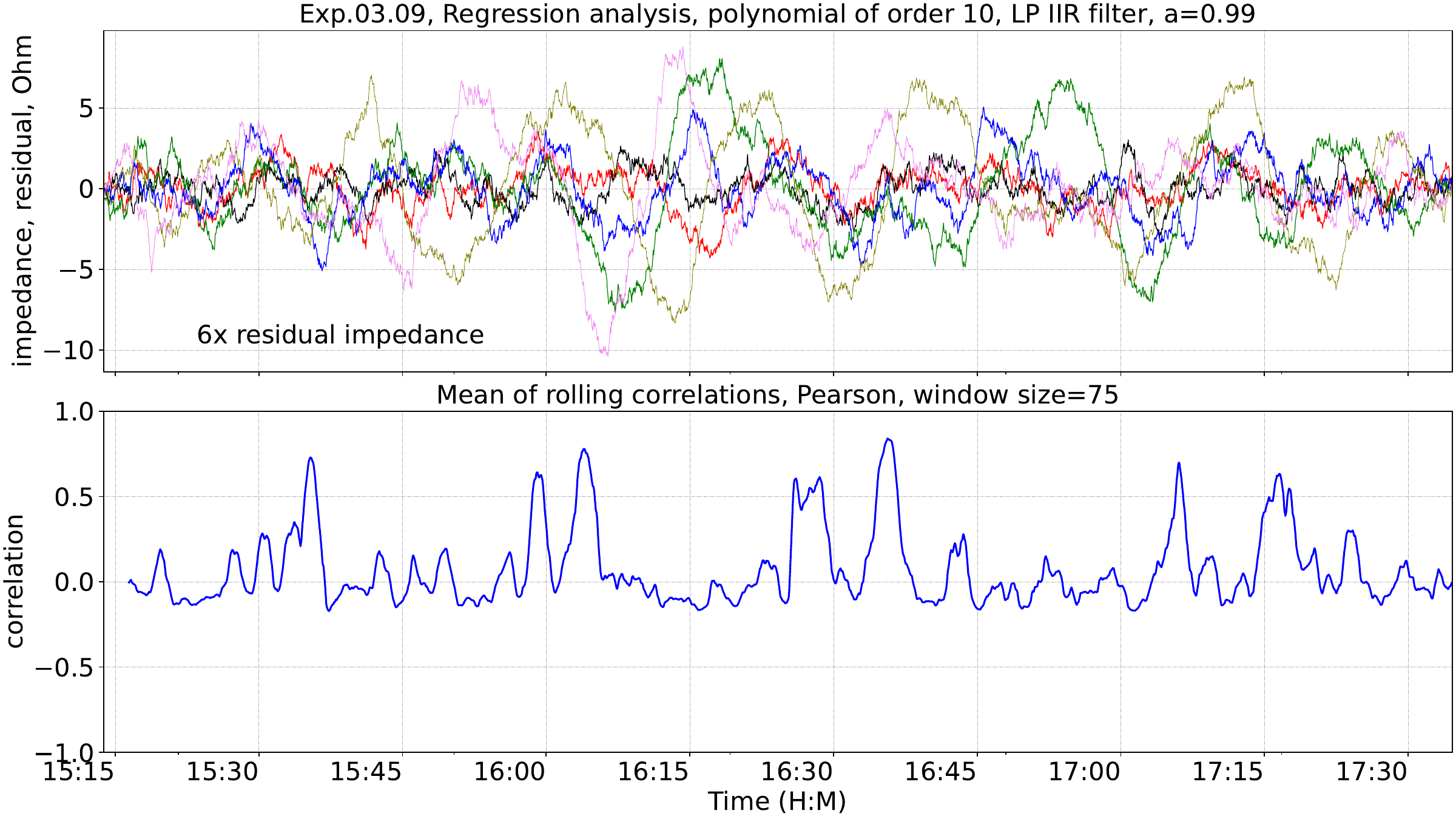}}
\caption{\textbf{(upper)} Residual impedance dynamics of six cells in optically non-transparent setup, we observe about 15 minutes difference in peaks; \textbf{(lower)} Mean of rolling correlations. \label{fig:exp0309_desynchroniz}}
\end{figure}

Optically non-transparent setups, such as shown in  Fig. \ref{fig:lightIsolatedSetup} or setups with several thermo-insulated boxes shown in Fig. \ref{fig:CO2_scenario_B}, generate individual electrochemical waves in each cell similarly to any other experiments. However, these waves are frequently not synchronized with each other, see Fig. \ref{fig:exp0309_desynchroniz} (and Fig. \ref{fig:oscillatingReactionsAdd} with a high rolling mean correlation). 

\begin{figure}[htp]
\centering
\subfigure{\includegraphics[width=0.49\textwidth]{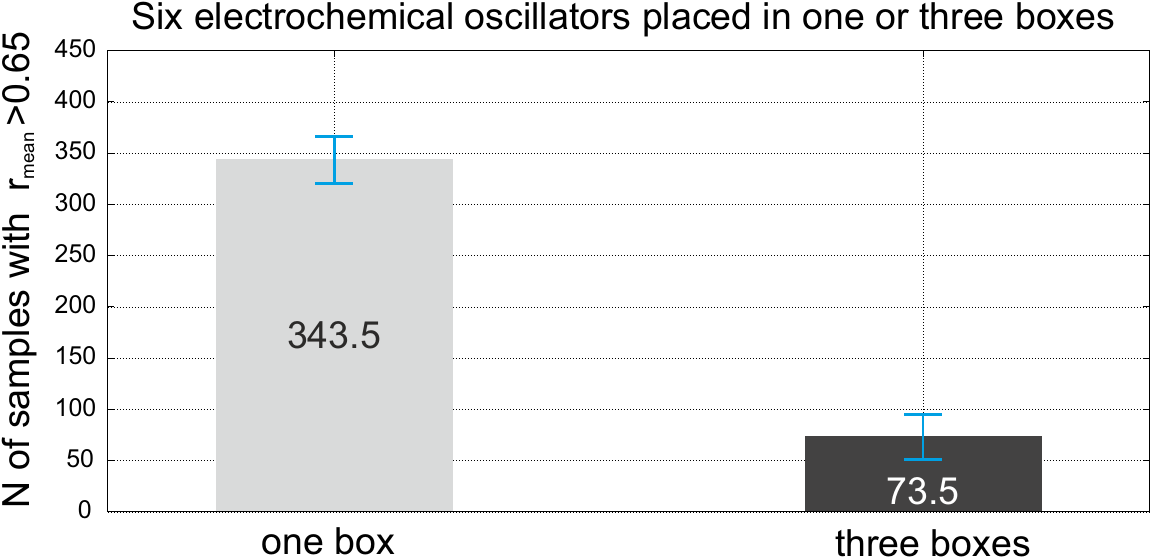}}
\caption{Six electrochemical oscillators placed in one (close to each other) or three different thermostabilizing boxes. Shown are $N$ samples per experiment with mean of rolling correlations $r_{mean}>0.65$, calculated in all experimental series in \ce{CO_2} scenario with totally 7.6 millions of samples (545181 data sets from 14 oscillators). Difference between these two cases is about 4.67 times, size of the correlation window is 100. \label{fig:one_three_boxes}}
\end{figure}

Comparing setups with oscillators placed in one thermo-insulating container and in several containers, we observe more synchronization events in the first case. Typically, 3-4 days experiment produces about 200000 data sets from 14 or 16 oscillators resulting in 2.8-3.2 millions of impedance-temperature samples. We analyzed 7.6 millions of samples from several experiments to find synchronization events in two cases: when oscillators are placed close to other in one thermo-insulating container or in different thermo-insulating containers (within the same experiment). The first case produces about 343 correlated data sets per experiment, the second case -- about 73 correlated data sets per experiment, see Fig. \ref{fig:one_three_boxes}. Thus, optically transparent setups generate about 4.67 times more synchronization events.

\subsection{Appearance of temperature-impedance waves}
\label{sec:TempImpWaves}

\begin{figure}[htp]
\centering
\subfigure[]{\includegraphics[width=0.49\textwidth]{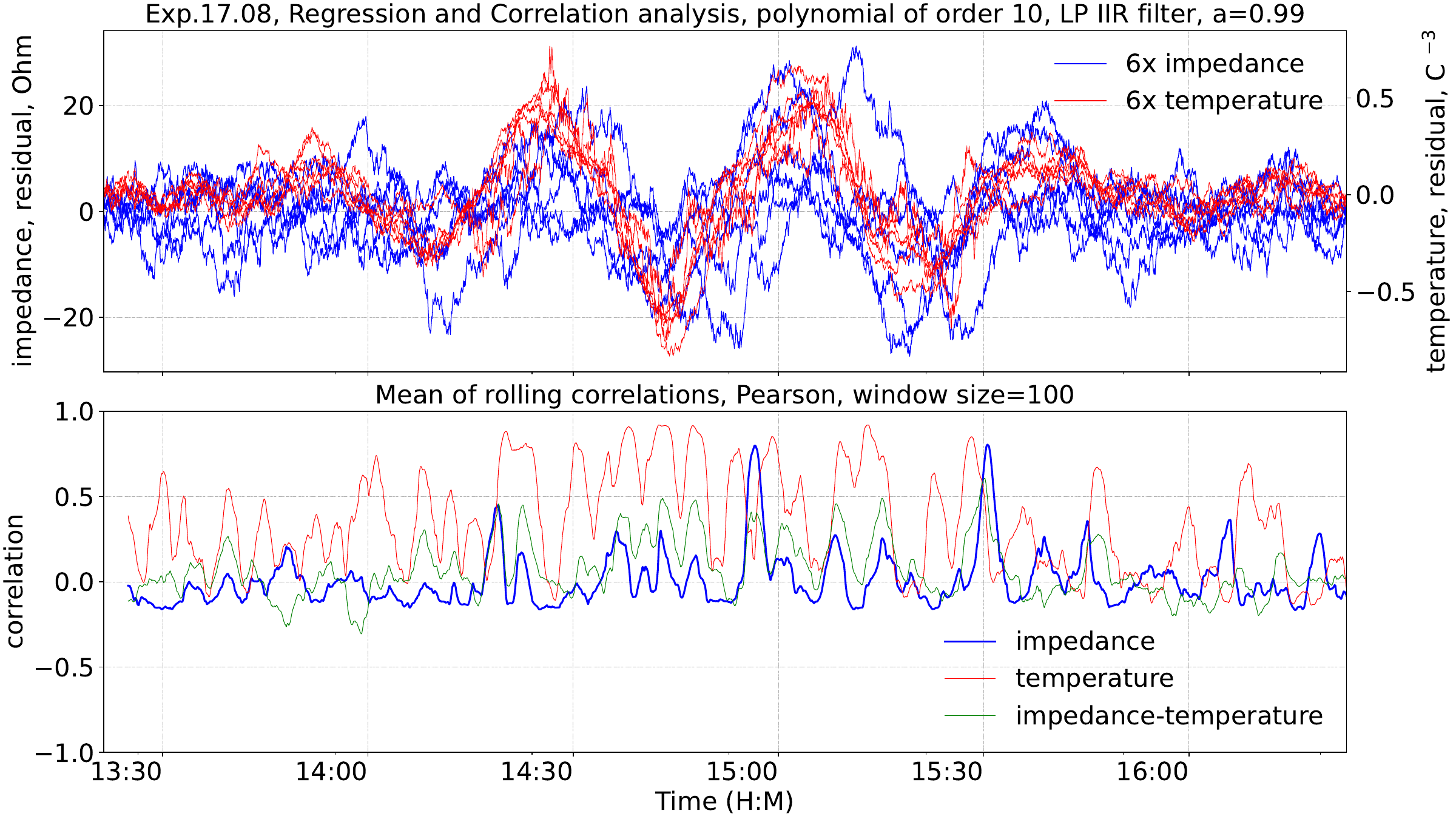}}
\subfigure[\label{fig:oscillatingReactions_temp}]{\includegraphics[width=0.49\textwidth]{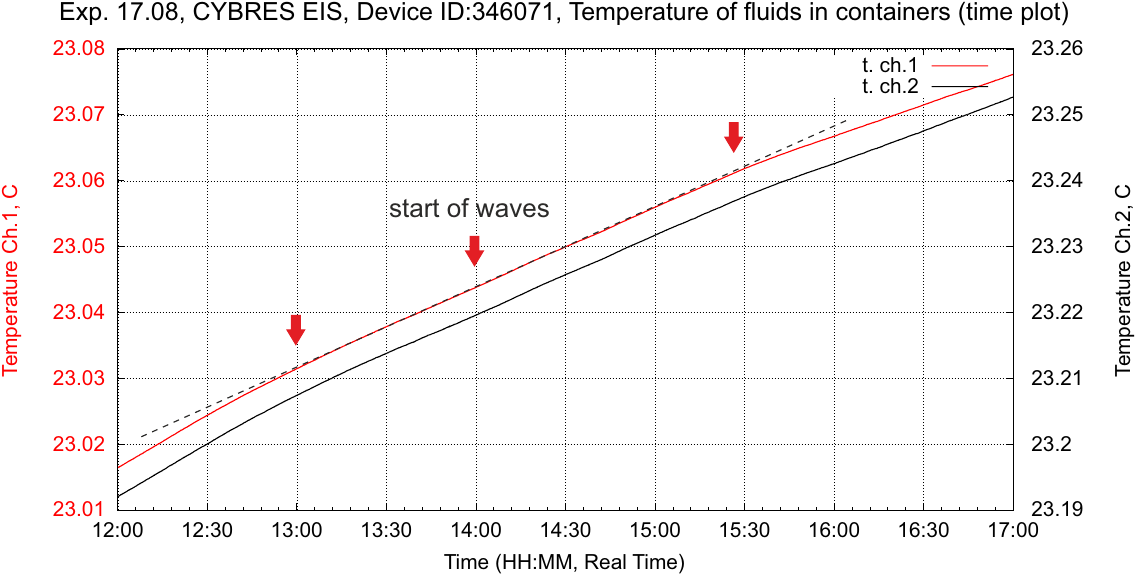}}
\caption{Example of temperature-impedance waves with the same phase. \textbf{(a)} Residual impedance and temperature dynamics of six cells together with means of rolling correlations in three groups; \textbf{(b)} Temperature of two fluidic cells in this experiment, temperature-impedance waves begin in a region of temperature with the stable trend. 
\label{fig:oscillatingReactions2}}
\end{figure}

Experiments demonstrated not only different electrochemical reactivity of isomers but also their different heat capacity \cite{kernbach23Thermal}, surface tension and capillary forces \cite{kernbach2023Pershin}, evaporation \cite{POULOSE2023814} and several other parameters. Thus, considering a spin-based character of EIS changes, we expect to discover oscillations of thermal parameters, evaporation rate and surface tension. Conducting experiments, we noted an interesting effects of generating short-term and low-amplitude temperature-impedance waves. Fig. \ref{fig:oscillatingReactions2} shows one example of such waves that begin in a region of temperature with the stable trend, see Fig. \ref{fig:oscillatingReactions_temp}.   

\begin{figure}[htp]
\centering
\subfigure{\includegraphics[width=0.49\textwidth]{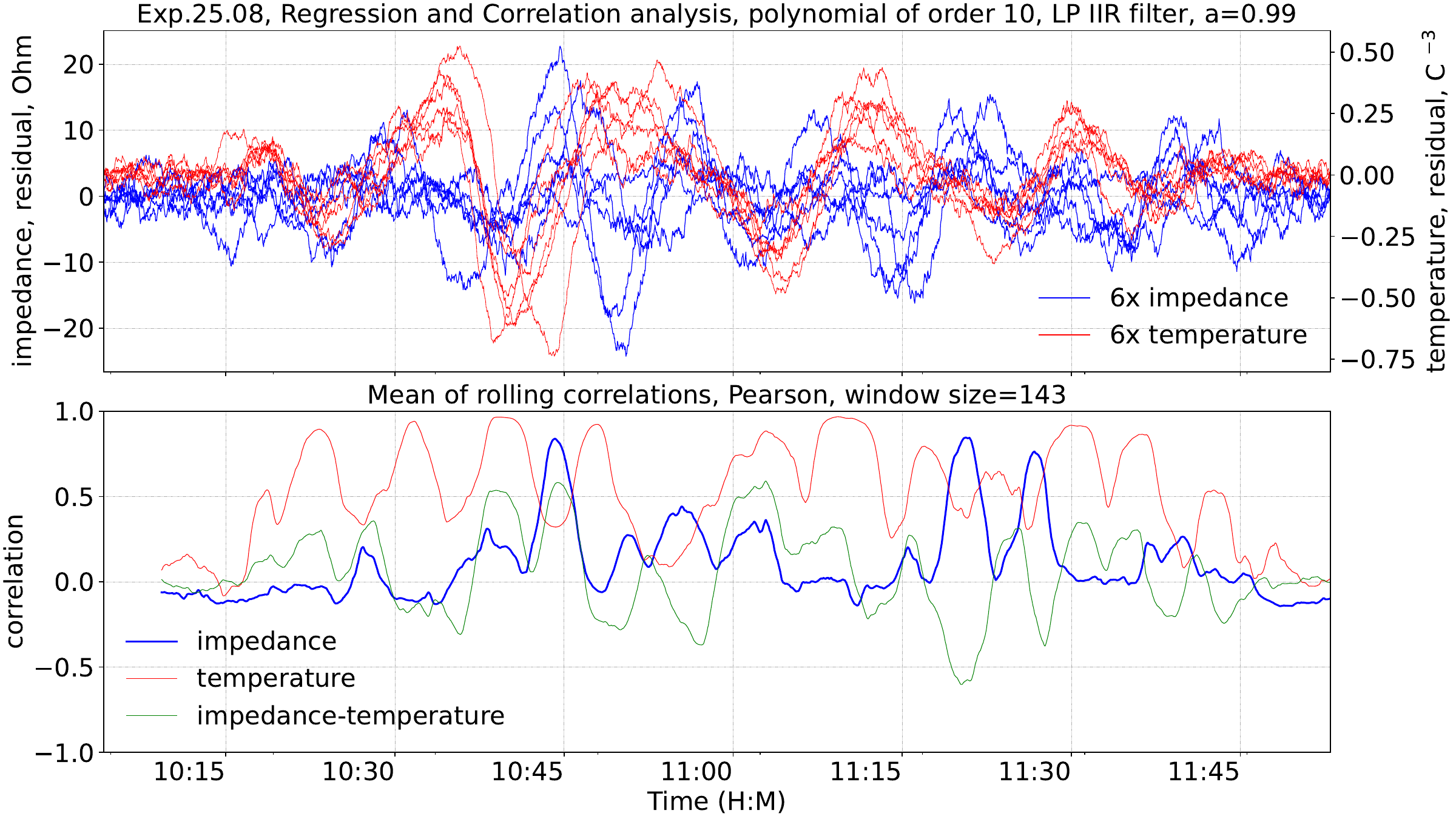}}
\caption{Example of temperature-impedance waves in the anti-phase. Shown are residual impedance and temperature dynamics of six cells together with means of rolling correlations in three groups; peaks of temperature-impedance correlations are negative. \label{fig:oscillatingReactions3}}
\end{figure}

Such temperature-impedance waves can have positive in-phase correlations, as shown in Fig.\ref{fig:oscillatingReactions2}, or negative anti-phase correlations, see Fig. \ref{fig:oscillatingReactions3}. Such a dynamics is not described by (\ref{eq:t}) and indicates their independent nature. These waves can take up to 5-6 oscillations with duration up to 90 minutes. We assume here the same spin conversion mechanism with  oscillating heat capacity as described in \cite{kernbach23Thermal}.

\section{Conclusion}

Performed experiments demonstrated a synchronization of molecular oscillators observable on a macroscopic scale that can be associated with the correlated switching between forward/reverse ionic production during \ce{CO_2} dissolving. Control attempts without \ce{CO_2} input demonstrated considerably lower number or even no such effects. Synchronization events typically take about 3-10 minutes and cannot be related to variations of \ce{CO_2} level or environmental temperature. Other environmental parameters such as a high-frequency electromagnetic emission or magnetic fields can potentially affect ionic production, but their monitoring did not reveal these factors during experiments. 

Electrochemical dynamics in one fluidic cell is characterized by multiple uncoordinated reactions generating chaotic macroscopic dynamics. Synchronization of molecular oscillators in one fluidic cell includes molecular coupling mechanisms; EIS dynamics in such cases demonstrates high-amplitude waves indicating more (in case of forward reactions) or less (in case of reverse reactions) ionic products in the fluid. This is an ongoing process taking place even at a low \ce{CO_2} input and/or along other reactive pathways such as dissolved oxygen and ROS reactions -- it is well measurable by EIS equipment. Since the mutual synchronization event represents a single wave, see Fig. \ref{fig:singleWave}, reflecting the coordinated reactions (\ref{eq:carbonicAcid})-(\ref{eq:carbonicAcidIonsFurther}), we reject the null-hypothesis about a false-positive interpretation of synchronization as a random overlapping of individual oscillations.

Synchronization of such waves between different cells attracted attention since their coupling mechanism cannot be explained by molecular interactions. Synchronization events between two and four cells have a high probability and show correlations with $r>0.9$. Increasing the number of cells decreases the probability of mutual synchronization; events in $\leq$10 cells with $r>0.7$ are still measurable. Appearance patterns indicate a random character of single events (once per 3000-8000 samples) and prevalence of anti-phase correlations. Separation of cells between different thermo-insulating non-transparent boxes (like 4+4+4 cells) reduces the number of synchronization events by the factor 4-5 and proposes light-matter interactions for a part of the cell-cell coupling mechanism. Considering a quantum nature of spin-conversion process, we assume the photon-assisted entanglement \cite{Zhang20}, which has been already demonstrated in a number of different quantum oscillators. Generation of in-phase and anti-phase correlated thermal and electrochemical waves confirms the hypothesis of spin-based effects since the isomers have not only different ionic reactivity but also heat capacity \cite{kernbach23Thermal}. We can also expect here oscillations of surface tension and evaporation rate \cite{Morre16, Drozdov14}.  

The coupling mechanism in cases of non-transparent setups is unclear. This is also related to the triggering event for synchronizations since we did not observe any obvious patterns for their appearance. Several theories, e.g. the pilot wave theory \cite{Trukhanova23} or spin-axion interactions can be considered for this mechanism. For future works, we see several possibilities, e.g. treating biophotonic effects \cite{POPP02} in term of quantum biology; low-cost spin sensors based on EIS or mobile MNR; exploring computations in macroscopic quantum networks.

\section{Acknowledgement}

This work is partially supported by EU-H2020 Project 'WATCHPLANT: Smart Biohybrid Phyto-Organisms for Environmental In Situ Monitoring', grant No: 101017899 funded by European Commission. Author thanks S.M.Pershin, M.Trukhanova and V.Zhigalov for fruitful discussions about spin conversion mechanisms, spin-spin and spin-axion interactions, and spin-based sensors.

\small

\appendix
\section{Supplementary information}

\begin{figure}[htp]
\centering
\subfigure{\includegraphics[width=0.45\textwidth]{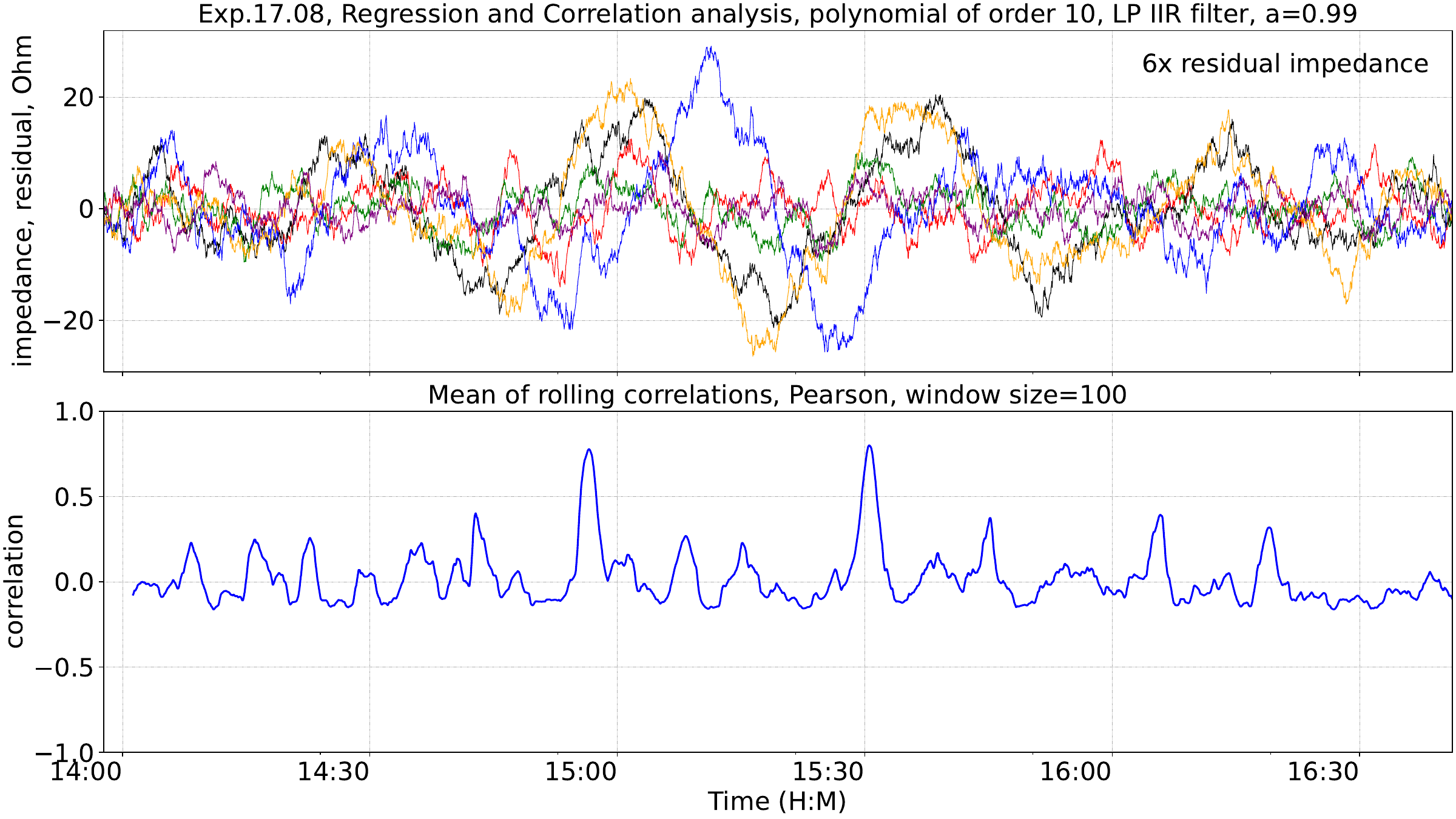}}
\caption{\textbf{Supplementary information to Sec. \ref{sec:severalContainers} }: Desynchronization of electrochemical dynamics in six cells that still demonstrate high rolling mean correlations due to sliding window. \label{fig:oscillatingReactionsAdd}}
\end{figure}

\begin{figure}[htp]
\centering
\subfigure[]{\includegraphics[width=0.45\textwidth]{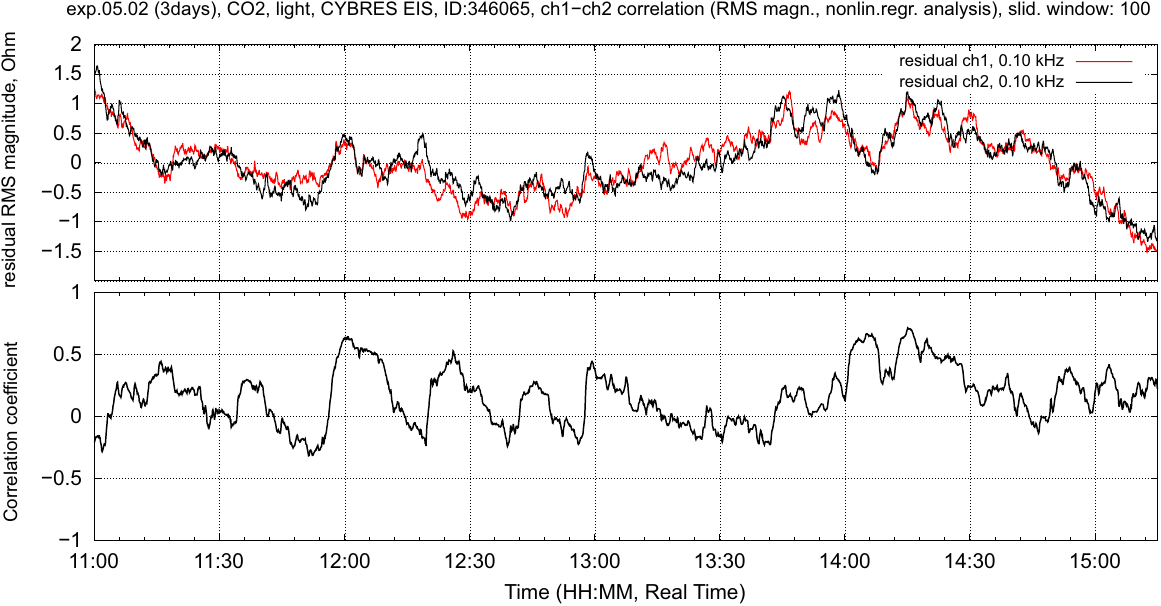}}
\subfigure[]{\includegraphics[width=0.45\textwidth]{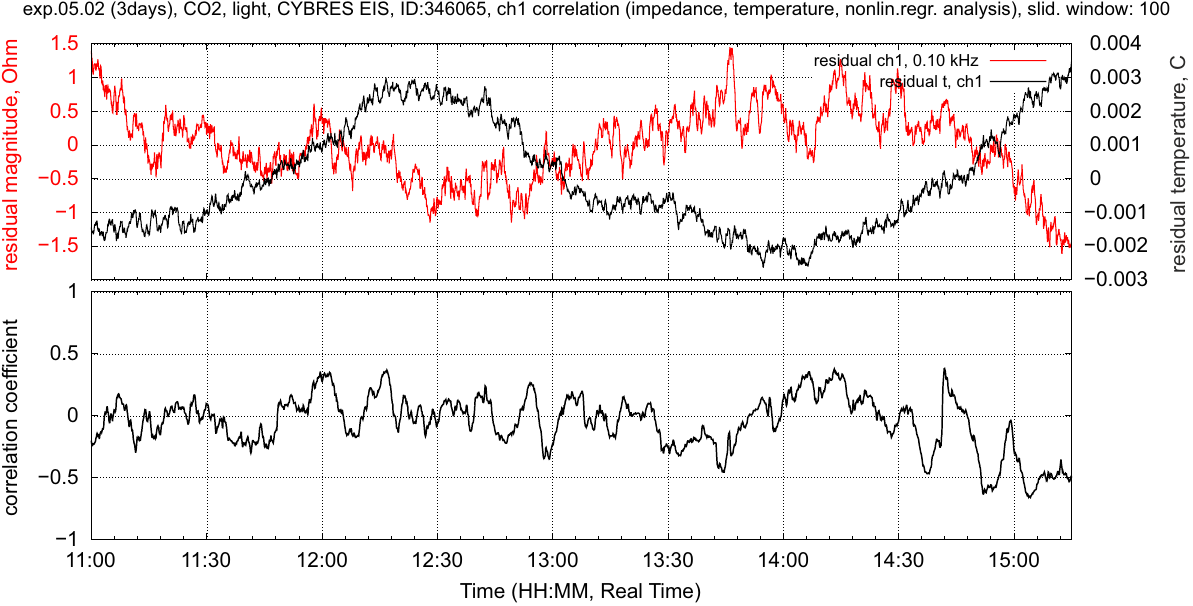}}
\caption{\textbf{Supplementary information to Sec. \ref{sec:controlTemperature}}: \textbf{(a)} Residual impedance-impedance dynamics of two cells -- multiple correlations are well visible; \textbf{(b)} Residual temperature-impedance dynamics for the same time period as in (a) -- beside macroscopic anti-phase dynamics of impedances and temperature, no short-term correlations are observed. \label{fig:exp050222Add}}
\end{figure}

\begin{figure}
\centering
\subfigure[\label{fig:shortTermCorr1}]{\includegraphics[width=0.49\textwidth]{./images/exp130222EIS99_1}}
\subfigure[]{\includegraphics[width=0.49\textwidth]{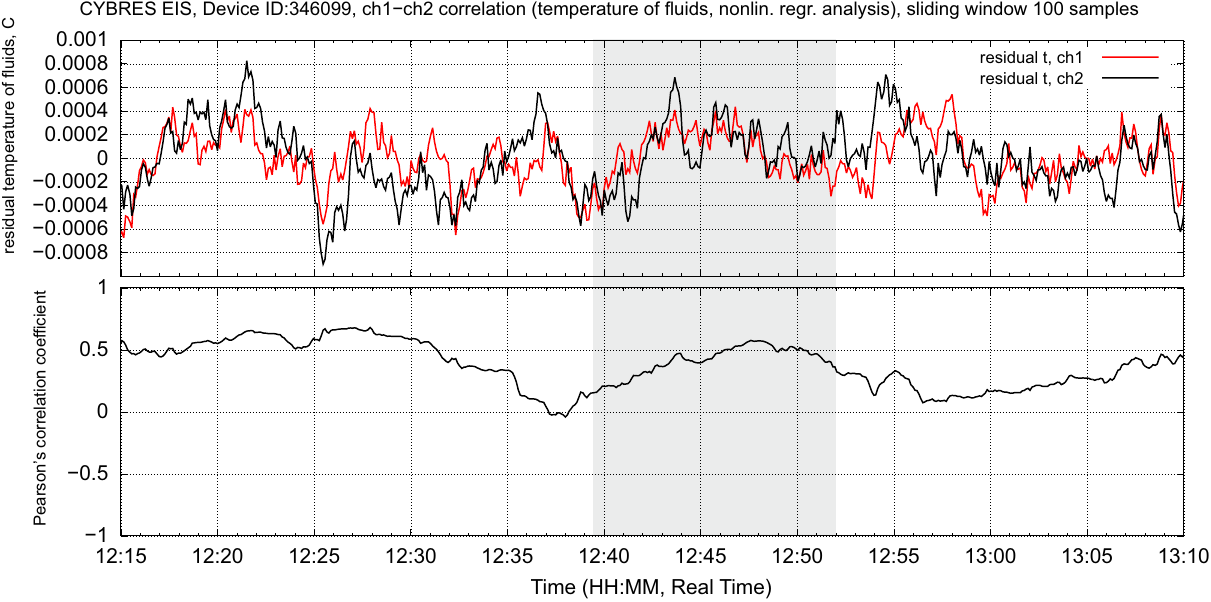}}
\subfigure[]{\includegraphics[width=0.49\textwidth]{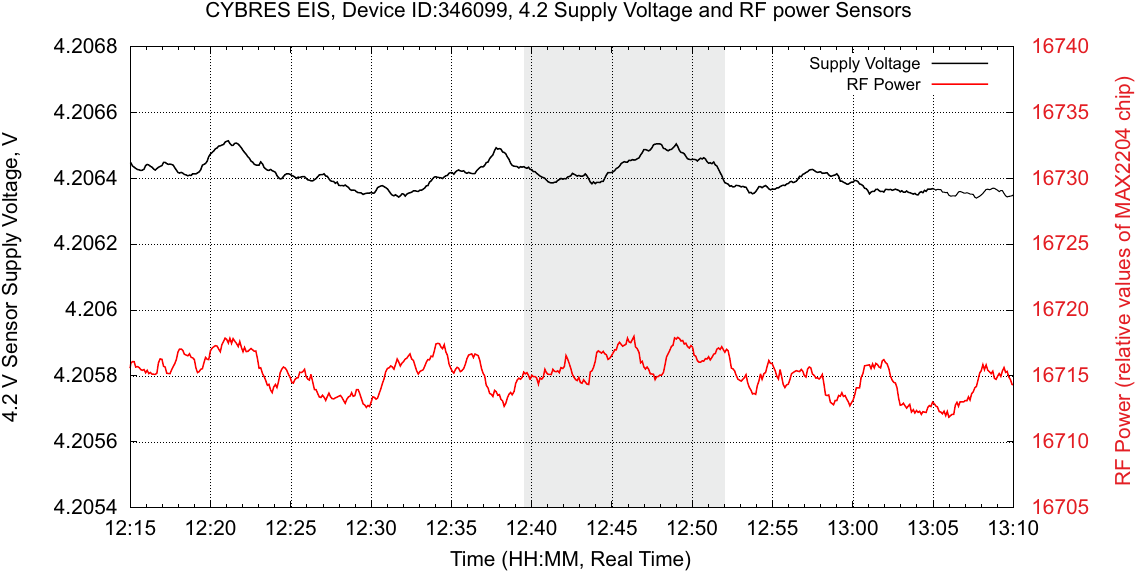}}
\subfigure[]{\includegraphics[width=0.49\textwidth]{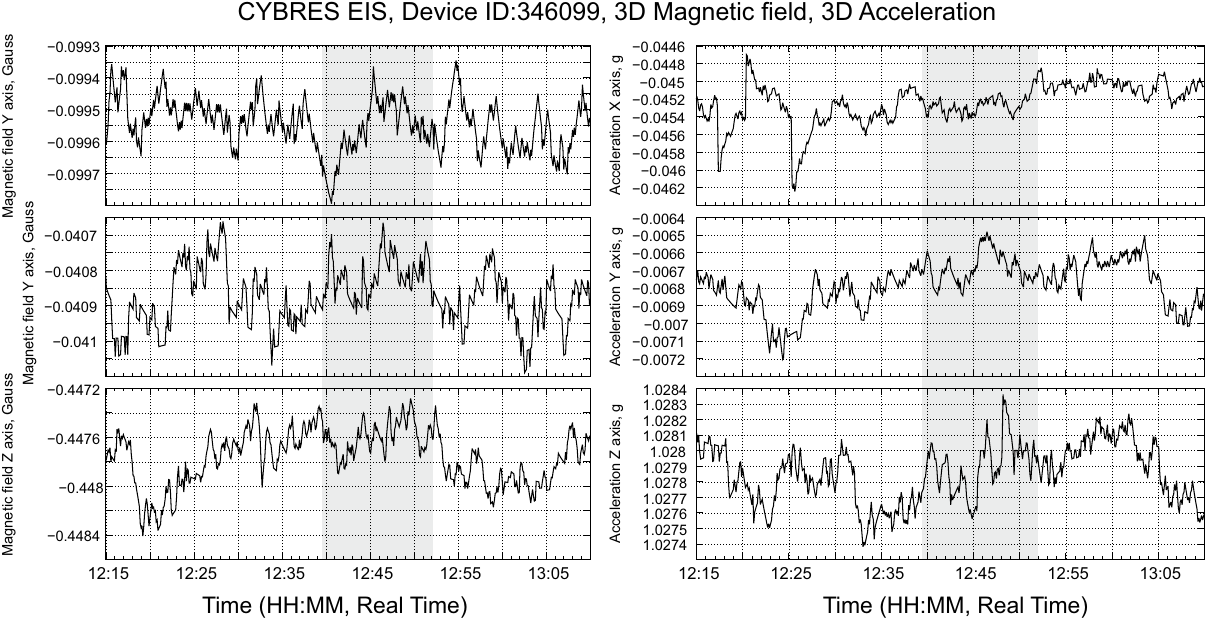}}
\caption{{\bf Supplementary information to Fig. \ref{fig:shortTermCorr}: (a)} Example of a short-term correlated dynamics of electrochemical impedances;  Non-correlated dynamics of {\bf (b)} fluid temperatures, {\bf (c)} power supply/RF power sensor, {\bf (d)} 3D accelerometer/magnetometer. \label{fig:shortTermCorrAdd}}
\end{figure}

\end{document}